\documentclass[a4paper,11pt]{article}
\pdfoutput=1
\usepackage{jcappub}
\usepackage[T1]{fontenc}
\usepackage{graphicx}
\usepackage{amsmath,amssymb}
\usepackage{hyperref}
\usepackage{url}
\usepackage{epstopdf}
\allowdisplaybreaks
\numberwithin{equation}{section}
\usepackage{enumitem}
\usepackage{bm}
\usepackage[utf8]{inputenc}
\usepackage{color}
\usepackage{bm}
\usepackage{multirow}
\usepackage{hyperref}

\usepackage{subfigure}
\usepackage{array}
\usepackage{todonotes}
\usepackage[english]{babel}
\usepackage{tensor}
\usepackage{comment}
\usepackage[normalem]{ulem}

\hypersetup{
    colorlinks=true,
    linkcolor=blue,
   citecolor=blue,
   urlcolor=blue
}

\graphicspath{{Figures/}}

\definecolor{nicered}{rgb}{0.7,0.1,0.1}
\definecolor{nicegreen}{rgb}{0.1,0.5,0.1}

\definecolor{vdrgreen}{rgb}{0.0, 0.7, 0.0}

\definecolor{abpurple}{rgb}{0.75, 0.0, 1.0}

\definecolor{blue(ncs)}{rgb}{0.0, 0.53, 0.74}

\usepackage{orcidlink}

\title{{\tt NAJADS}: a self-contained framework for the direct determination of astrophysical J-factors}

\author[a,b]{Anna Balaudo\orcidlink{0000-0003-4109-8094},}
\emailAdd{balaudo@strw.leidenuniv.nl}
\author[c]{Francesca Calore\orcidlink{0000-0001-7722-6145},}
\emailAdd{calore@lapth.cnrs.fr}
\author[d]{Valentina De Romeri\orcidlink{0000-0003-3585-7437},}
\emailAdd{deromeri@ific.uv.es}
\author[e,f]{Fiorenza Donato\orcidlink{0000-0002-3754-3960}}
\emailAdd{donato@to.infn.it}

\affiliation[a]{Leiden Observatory, Leiden University, PO Box 9513, Leiden 2300 RA, The Netherlands}
\affiliation[b]{Institute Lorentz, Leiden University, PO Box 9506, Leiden 2300 RA, The Netherlands}
\affiliation[c]{Laboratoire d'Annecy-le-Vieux de Physique Théorique (LAPTh), CNRS, USMB, 9 Chemin de Bellevue, 74940 Annecy, France}
\affiliation[d]{Institut de F\'{i}sica Corpuscular CSIC/Universitat de Val\`{e}ncia, Parc Cient\'ific de Paterna C/ Catedr\'atico Jos\'e Beltr\'an, 2 E-46980 Paterna (Valencia), Spain}
\affiliation[e]{Department of Physics, University of Torino, via P. Giuria, 1, 10125
Torino, Italy}
\affiliation[f]{Istituto Nazionale di Fisica Nucleare, Sezione di Torino, via P. Giuria, 1, I-10125 Torino, Italy}

\abstract{
Cosmological simulations play a pivotal role in understanding the properties of the dark matter (DM) distribution in both galactic and galaxy-cluster environments.
The characterization of DM structures is crucial for informing indirect DM searches, aiming at the detection of the annihilation (or decay) products of DM particles.
A fundamental quantity in these analyses is the astrophysical J-factor. In the DM phenomenology community, J-factors are typically computed through the semi-analytical modelling of the DM mass distribution, which is affected by large uncertainties.
With the scope of addressing and possibly reducing these uncertainties, we present {\tt NAJADS}, a self-contained framework to derive the DM J-factor directly from the raw simulations data.
We show how this framework can be used to compute all-sky maps of the J-factor, automatically accounting for the complex 3D structure of the simulated halos and for the boosting of the signal due to the density fluctuations along the line of sight.
After validating our code, we present a proof-of-concept application of {\tt NAJADS} to a realistic halo from the IllustrisTNG suite, and exploit it to make a thorough comparison between our numerical approach and traditional semi-analytical methods.
}

\begin{document}

\keywords{astroparticle physics, cosmological simulations, dark matter indirect detection, J-factor}
\maketitle

\section{Introduction}
\label{sec:intro}

Probing the nature of the most elusive matter component in the universe, 
the dark matter (DM), is one of the major endeavours of physics, 
sitting at the intersection between particle physics, astrophysics and cosmology.
Astroparticle observables can be very successful in testing the DM nature for a broad range of candidates~\cite{Batista:2021ctk}.
In particular, in the absence of positive detection so far, gamma-ray searches for DM signals from MeV up to TeV energies provide us with the strongest constraints to date on thermally produced particle DM~\cite{PerezdelosHeros:2020qyt, Taoso:2021gvk}.
These constraints are nonetheless affected by large systematic uncertainties related mostly to the DM 
distribution in the target of interest. Examples are the impact of the DM mass modelling of the Galactic halo at high latitudes~\cite{Eckner:2022swf}, and 
in dwarf spheroidal galaxies~\cite{Ullio:2016kvy}.
Thus, one crucial ingredient for indirect DM searches is the DM distribution in astrophysical systems, from dwarf-sized galaxies to large galaxy clusters. 
More specifically, the expected photon flux from DM annihilation or decay is proportional to the integral along the line of sight (LOS) of the DM spatial distribution, eventually squared in the case of DM particle annihilation. 
For example, for a generic self-conjugated DM candidate, the expected gamma-ray flux, per unit energy and solid angle, writes as: 
\begin{equation}
    \label{eq:annihilation_flux}
    \frac{d\Phi}{dEd\Omega} = \frac{1}{4 \pi} \frac{\langle \sigma v \rangle}{2 m^2} \frac{dN}{dE}{}\int_{\rm LOS} \rho^2 (l(\psi) ) dl(\psi) \, ,
\end{equation}
where $\langle \sigma v \rangle$ is the annihilation cross section averaged over the DM particles  velocity distribution, and $m$ is the mass of the DM particle candidate. 
The integral along the LOS of the DM density squared is dubbed J-factor (and, similarly, D-factor for decaying DM). 
We can therefore define:
\begin{equation}
		\label{eq:jfactor_definition}
			J = \int_{\rm LOS} \rho^2 (l(\psi) ) dl(\psi) \, .
		\end{equation}
Here $\psi$ is the angle between the direction of the halo centre and that
of observation.

From Eq.~\eqref{eq:annihilation_flux}, we can immediately realise that, especially for annihilation, 
the ensuing signal can be very sensitive to the uncertainties in the DM density profile, $\rho$.
As observers on Earth, we indeed receive the flux from the total DM distribution in the Galaxy (and beyond). 
Besides the flux from the smooth Milky-Way halo, the presence of substructures further affects the predictions for DM signals.
On the one hand, unresolved DM over-densities on the top of the smooth DM halos enhance the DM expected flux by the so-called "boost" factor, see~\cite{Ando:2019xlm} for a review. On the other hand, one can target the signal directly originating from resolved DM sub-halos, as individual point-like or extended sources, for example looking for unidentified sources of X- and gamma-ray radiation (see e.g.~\citep{Bertoni:2015mla,Schoonenberg:2016aml,Hooper:2016cld,Calore:2016ogv,Coronado-Blazquez:2019pny,Coronado-Blazquez:2019puc,Coronado-Blazquez:2019bkv,DiMauro:2020uos,HESS:2021pgk}),
or for anisotropies in cosmic radiation backgrounds~\citep{Siegal-Gaskins:2008usr,Ando:2009fp,Fornasa:2012gu,Fermi-LAT:2012pez,Calore:2014hna,Hutten:2018wop}.

In the presence of sub-halos, the total J-factor is the sum of different components, taking into account also cross-correlation terms, cf.~Eqs.~(13)--(16) in~\cite{2012CoPhC.183..656C}.
Given the large number of sub-halos in galaxies, computing the total J-factor for a realistic DM distribution is numerically challenging, and semi-analytical codes like {\tt CLUMPY}~\cite{2012CoPhC.183..656C,Hutten:2018aix} have been developed for this purpose, additionally adopting some statistical approach to account for low-mass sub-halos.

Besides numerical challenges, there are also modelling ones. 
Indeed, the J-factor computation must rely on theoretical models for the matter distribution in the main halo and sub-halos.
The DM distribution in galaxies can be constrained by observational data, but its derivation is still prone to large uncertainties, especially for the inner regions of the Milky Way and ultra-faint dwarf galaxies (see e.g.~\cite{Pato:2015tja,Bonnivard:2015xpq,Chang:2020rem}).
To get an estimate of systematic uncertainties to be used for indirect signal predictions, the DM phenomenology community typically relies on semi-analytical models of the halo statistics, e.g.~\cite{Stref:2016uzb,Ando:2019xlm}, or on the outcome of numerical simulations for galaxy formation, which provide a proxy for the DM distribution at different scales, as well as a way to quantify statistical and systematic uncertainties by analysing large samples of objects (e.g.~\cite{Calore:2016ogv,Coronado-Blazquez:2019puc,Grand:2020bhk}).

In the past decades, a huge effort has been put into finding a universal parametrisation to describe the density profiles of DM halos. These parametrisations are typically based on the assumption of spherical symmetry and contain a number of free parameters that need to be properly tuned to match the shape of peculiar halos. Arguably, the most famous density profiles present in the literature include central cored profiles, of which the Burkert parametrisation can be considered the main representative \citep{Burkert:1995yz}, along with a family of cuspy profiles, such as the Einasto \citep{Einasto:1965czb,Graham:2005xx,Navarro:2003ew} and the Navarro, Frenk and White parametrisation (hereafter NFW) \citep{Navarro:1995iw, Navarro:1996gj} with its many generalizations, the Moore \citep{Moore:1999gc}, Hernquist \citep{Hernquist:1990be} and the Dehnen $\&$ McLaughlin profiles \citep{Dehnen:2005cu} and many others. While all these functions present quite similar behaviours at high radii, they provide significantly different predictions on the DM density when extrapolated towards the centre of the halo.
Plus, in contrast with the common phenomenological practice of describing the matter distribution with spherical profiles, it is often found that in DM-only simulations halos tend to have a prolate shape, and the most massive ones can even present a triaxial structure (see e.g. \citep{Despali:2014via}). In hydrodynamical simulations the shapes of halos, notably Milky-Way sized objects, are predicted to be more spherical~\citep{2019MNRAS.484..476C, 2019MNRAS.490.4877P}. Nonetheless, a mild, possibly non-negligible deviation from sphericity remains.

Even larger uncertainties affect the determination of the DM distribution in sub-halos.
Typically, it is assumed to follow the same parametrisation of 
the parent halo, although past works showed that halo shapes can depend significantly on the halo mass~\citep{Gao:2007gh}.
Additionally, sub-halos go through remarkably different evolution histories than their field host, and may suffer heavy tidal stripping
which strongly affects their DM distribution, see e.g.~\citep{Diemand:2007qr,vandenBosch:2017ynq,vandenBosch:2018tyt,Green:2019zkz}.

Determining the halos and sub-halos abundance, mass and spatial distribution is
typically done with the aid of halo finders, which are algorithms applied
to numerical simulations of structure formation to identify over-dense regions and to extract some of their main properties (for comprehensive reviews see for example~\cite{Knebe:2011rx,Onions:2012mh,Knebe:2013xwz}).
These algorithms scan the simulation box, oftentimes exploiting the full 6-dimensional (spatial and velocity) information, looking for local peaks in the density field and identifying samples of gravitationally bounded particles, which then define the (sub-)halos. 
Among the most commonly adopted halo finders, we recall, for example, Friends of Friends ({\tt FoF})~\citep{Davis:1985rj,Knebe:2011rx}, {\tt SUBFIND}~\citep{Springel:2000qu}, Amiga Halo Finder ({\tt AHF})~\citep{Knollmann:2009pb}, but also {\tt VOBOZ}~\citep{Neyrinck:2004gj}, {\tt 6DFOF}~\citep{Diemand:2006ey}, {\tt ROCKSTAR}~\citep{Behroozi:2011ju} and {\tt VELOCIRAPTOR}~\citep{Elahi:2019wap}.
Such a delicate procedure of identifying substructures in the simulations necessarily relies on some assumptions.
First, the mass definition of bounded over-densities is not universal among different algorithms. This is a crucial issue, since whether a particle is  bound to a structure depends on the mass of the object itself. Moreover, the mass estimate also affects the derivation of the analytical DM distribution parameters.
Comparisons of halo finders performances showed that the mass estimates for the same sub-halo can differ up to $20-30\%$ for most massive structures, while differences can reach an order of magnitude or more for less massive objects~\citep{Muldrew:2010hx,Knebe:2011rx,Knebe:2013xwz}.
Secondly, there are no universal ways of defining the edge of sub-structures, nor recipes for treating those outskirts transition regions either as contributing to the properties of the host halo or as completely separated objects.
Lastly, even if different halo finders identify the same set of particles for a halo, there exist different possibilities for how to derive its properties, depending on the assumed initial conditions and on the mass/edge definitions previously mentioned.

All these model uncertainties will severely affect subsequent phenomenological applications, such as the final estimate of the J-factor and the resulting astrophysical constraints on DM properties.
With the final goal of addressing and possibly mitigating some of these systematics, we build the present work on the following question:
can we bypass the use of halo finders and derive the J-factor directly from simulations' raw data? 
While the simulation community already adopts these more intrinsic methods to compute the density field we aim here at systematically compare them with the analytical predictions, instead based on halo finders. We focus on the impact on dark matter J-factors, which at the very end affects the expected fluxes of observable particles.
Although we fully acknowledge that halo finders provide an extremely powerful tool to analyse sub-halo populations properties and characterising their evolution through cosmic time, they may not be an indispensable tool to extract predictions of DM indirect signals. 
Specifically to this end, we design {\tt NAJADS} (Numerical Astrophysical J-factors for Annihilating Dark matter Signatures), a code that proposes a first, seminal step towards the accomplishment of such a delicate task\footnote{{\tt NAJADS} is not publicly available at the present time. We plan on this to change in the next future.}. 
Let us clarify here that our goal is not to produce a novel, cutting-edge density estimator to improve upon existing tools, but rather to implement such tools within a self-consistent, simple and complete framework, specifically designed for the numerical calculation of astrophysical J-factors. In this sense,
{\tt NAJADS} offers a free-standing infrastructure to go directly from the raw simulation’s output to the J-factor quantity, without requiring the analytical modelling of the DM distribution.
We notice that a similar approach is not unprecedented, and has been adopted recently, e.g., by the FIRE simulation~\cite{McKeown:2021sob}.
However, the implications for DM observables have not been fully explored nor a thorough comparison with traditional, analytical modelling methods has been carried out yet. We take advantage of {\tt NAJADS} to address both points in this work, and discuss further applications to DM astrophysical studies.

The paper is organised as follows.
In Sec.~\ref{sec:code_structure}, we outline the structure and core modules of the code, while in Sec.~\ref{sec:testing} we present some validation tests of {\tt NAJADS} on the Dragon catalogue, the Monte Carlo (MC) realisation of a mock cosmological simulation suite created for testing purposes. In Sec.~\ref{sec:proof_of_concept} we show an end-to-end application of {\tt NAJADS} through the evaluation of J-factors sky-maps directly from the mock simulations' raw data. As part of this proof of concept, we also run {\tt NAJADS} on a real cosmological simulation, selecting a halo from the IllustrisTNG suite\footnote{\url{https://www.tng-project.org/}}, comparing our framework with semi-analytical methods and discussing advantages and disadvantages of both.
Finally, we summarise our findings and discuss further developments in Sec.~\ref{sec:conclusions}. Moreover, we present in App.~\ref{app:mock_catalogue} the procedure followed to generate the Dragon suite, while further assessments of {\tt NAJADS} are discussed in App.~\ref{app:density_scatter}.

\section{The {\tt NAJADS} framework}
\label{sec:code_structure}

The code we developed for our exploration is organized in macro modules, performing different tasks. It is advantegeously built to allow each module to be replaced by a user-defined module of choice. This is purposefully intended to facilitate, for example, the deployment of a different algorithm for the density estimation, or of a different integration method than the default one that we propose here.

{\tt NAJADS} takes as input one or more DM halos, provided as an array of particle positions from the raw output of a cosmological simulation. With this, the algorithm can calculate J-factors either on individual, user defined, lines of sight (LOS), or it can produce an HEALPix~\citep{Gorski:2004by} skymap of the J-factor for a given halo, at any resolution - though, the higher the number of pixel, the more computation resources required. In both scenarios, the observer position can be defined by the user. {\tt NAJADS} is fully parallelized so that computations of J-factors on different lines of sight can occur simultaneously.

The first macro-module of {\tt NAJADS} is a nearest-neighbour algorithm to estimate the local density. As we decided not to use one of the many algorithms already available in the literature but to define our own, the details of the density calculations are briefly summarized below. The second macro-module is dedicated to the J-factor calculation along a LOS, implementing a discretization of Eq.~\eqref{eq:jfactor_definition}. The integration is performed through the trapezoidal rule, and the default version is customizable in that the user can control the integration step.

\subsection{Computation of the local density}
\label{subsec:local_density_estimate}

Our density calculation is based on the simple idea that, given a discrete distribution of point-like particles with a given mass, the density around each point can be computed associating to the particles an occupation volume $dV$. 
\\
In a uniform distribution, i.e.~a distribution in which the density is constant everywhere, each particle is ideally associated to a same $dV$, and the total volume of the simulation is given by the sum of all the infinitesimal volumes corresponding to different particles.
In a real-life situation, where the density is not uniform, even the localisation volumes will change accordingly, being smaller in regions where the density increases and larger in lower-density zones.

We assume the infinitesimal volume $dV$ to be equivalent to a cube of side $d$, thus $d^3 = dV$, where $d$ can  be regarded as an estimate of the average separation between a particle and its nearest neighbours. 
The determination of $dV$  allows the calculation of the local density as $\rho = m_p / d^3$, $m_p$ being the particle mass. Thus, the main steps of our density estimation algorithm are the following:

\begin{enumerate}
	\item Starting from the position $\bar{x}$ at which we wish to compute the density, we identify the positions $\bar{r}^n$, $n = (1,...,N)$, of the $N$ particles closest to $\bar{x}$.
	\item For each particle at $\bar{r}^n$, we compute the distance to the $j$ closest particles -- located at $\bar{r}^n_{ij}$ -- and average to find $d_j^n = \frac{1}{j} \sum_{i=1}^{j} |\bar{r}^n - \bar{r}^n_{ij}|$.
	\item Step (2) is repeated several times, varying $j$ from a minimum value $a$ to a maximum value $b$, then averaging again to obtain $d^n = \frac{1}{b-a}\sum_{j=a}^{b} d_j^n$.
	\item At this point,  we have an estimate of the mean distance $d^n$ between each $\bar{r}^n$ particle and its nearest neighbour, and we average one more time to find the mean virtual side in that simulation region: $d = \frac{1}{N}\sum_{n=1}^{N} d^n$.
	\item Finally we compute the density, $\rho(\bar{x}) = m_p/d^3(\bar{x})$.
\end{enumerate}
Eventually, the mass density at $\bar{x}$ reads:

\begin{equation}
\label{eq:density_estimate}
\rho(\bar{x}) = m_p \Bigg[\frac{1}{N}\sum_{n=1}^{N}\frac{1}{b-a}\sum_{j=a}^{b} \frac{1}{j}\sum_{i=1}^{j} |\bar{r}^n - \bar{r}^n_{ij}|\Bigg]^{-3} \,.
\end{equation}

A few comments on this definition are in order. First, in writing a suitable algorithm we introduced three separate averages, dependent on three parameters ($a$, $b$ and $N$) whose values can be chosen by the user. These averages stabilize {\tt NAJADS} response to small fluctuations in the particles positions, increasing the scale at which we measure density deviations. This scale is eventually fixed by the value of $N$ -- and marginally by $a$ and $b$ -- and defines the locality of the estimate: the smaller the value of $N$, the more local {\tt NAJADS} result will be. 
We found that values of $N$ that keep the density estimation sufficiently stable, but local enough to allow for the resolution of structures in cosmological simulations, range in the interval $[10,20]$. In the rest of our study we then restrict to $N = 15$. We however advise against using higher $N$ values, as while only minimally affecting estimates on the more resolved host halo, the loss of locality can have a huge impact in sub-halos density reconstruction.

The remaining couple $(a,b)$ is a measure of the amount of nearest neighbours around each particle. Specifically, it tells us that in the region under consideration particles can have a number of nearest neighbours that ranges from $a$ to $b$.
Very high values for $a$ and $b$ can eventually end up affecting the locality of the measure as well. Fixing these values \textit{a priori} from theoretical considerations is tricky, as the distance of a particle to the surrounding ones increases gradually, and there is no clear distinction between those particles that can be taken as nearest neighbours and those which are too far away to be considered for the density estimation. Hence, the couple $(a,b)$ needs to be fixed phenomenologically.
We recommend the choice $a=6$ and $b=13$, which are also the values that we adopted in the rest of this work. This choice is motivated in App. \ref{app:ab_determination} on a set of mock simulations.

The dependence of our density estimator on the three described hyperparameters is also its most significant limitation. We chose to implement our own version of a density estimator in the default setup of {\tt NAJADS} solely to rely on a simple but stable algorithm over which we had full control, so that we could interpret our results easily and consistently. We acknowledge that other density estimators are already available to the community, that possibly outperform our own in both speed and accuracy. We stress that it is not the intention of this work to produce a novel, cutting edge density estimator to compete with other tools, but rather, to provide a consistent and complete framework for the numerical calculation of astrophysical J-factors in DM halos. Specifically in light of this consideration, we designed {\tt NAJADS} in a modular way, so that it is already possible to call an external, alternative, modulus for the density calculation. We envision to further develop NAJADS in future works, expanding the code to include more options for the density estimation among the many available to the community, such as gaussian kernels methods, adaptive mesh grid algorithms and estimators exploiting the information on particle velocities in a full 6D analysis.

\section{{\tt NAJADS} to the test on mock cosmological halos}
\label{sec:testing}

\begin{figure}
	\centering
	\includegraphics[width=1.02\textwidth]{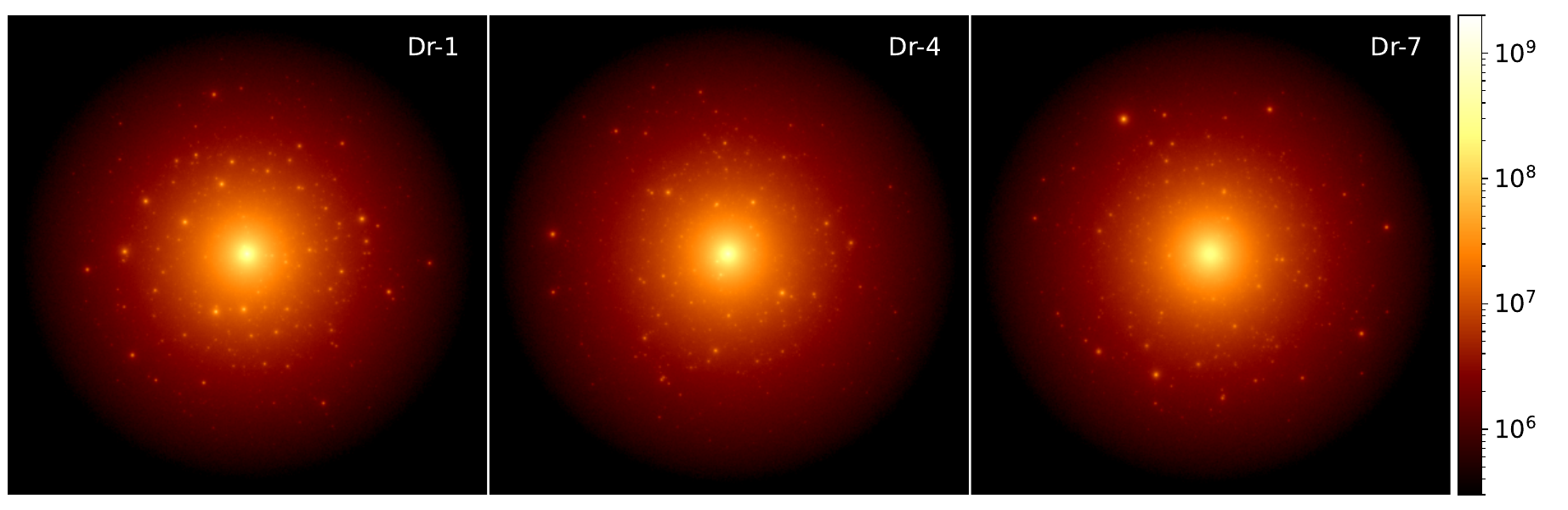}
	\caption{Surface density maps of three representatives mock halos, for an Einasto-like (Dr-1), NFW-like (Dr-4) and Burkert-like (Dr-7) radial mass distribution. Each box side measures 600 kpc. Units are $M_{\odot}/{\rm kpc}^2$.}
	\label{fig:mock_halos_surface_density}
\end{figure}

Before launching {\tt NAJADS} on real cosmological simulations, we produced for testing purposes a suite of nine MC realizations of galactic halos (hereafter, the Dragon catalogue) that mimic cosmological simulation outputs. Specifically, each of these halos is a 3D random realization of an injected analytical mass distribution. The details of the procedure we followed are presented in App.~\ref{app:mock_catalogue}.
We restrict our study to the Einasto (E) and the NFW cuspy density profiles, and to the cored Burkert (B) profile, all obeying a spherical mass distribution. Their parameterizations follow:

\begin{subequations}
	\label{eq:analytical_profile}
	\begin{align}
	& \quad \rho^{\rm E}(r) = \rho_{-2} \cdot {\rm exp}\left\{- \frac{2}{\alpha}\Big[ \Big(\frac{r}{r_{-2}} \Big) ^{\alpha}-1\Big]\right\} \, , \label{eq:analytical_profile_a} \\
	& \quad \rho^{\rm NFW}(r) = \frac{4 \rho_{-2}}{\Big(r/r_{-2}\Big)\Big(1+r/r_{-2}\Big)^2} \, , \label{eq:analytical_profile_b}\\
	& \quad \rho^{\rm B}(r) = \frac{4\rho_{-2}}{\Big(1+r/r_{-2}\Big)\Big[1+\Big(r/r_{-2}\Big)^2\Big]} \, .\label{eq:analytical_profile_c}
	\end{align}
\end{subequations}

Here, $r$ is the radial distance from the centre of the halo. The chosen analytical profiles have been usefully re-parametrised as done in \citep{Navarro:2008kc}, so that $r_{-2}$, $\rho_{-2}$ and $\alpha$ are free parameters representing, respectively: the radius at which the logarithmic slope of the profile equals the isothermal value, the corresponding density $\rho(r_{-2})$, and a slope parameter typical of the Einasto parametrisation that regulates the steepness of the profile. In what follows, we have kept $\alpha$ fixed at $0.17$ \citep{Springel:2008cc, Navarro:2003ew, Navarro:2008kc}. As for the resolution of our mock catalogue, each halo as a mass of $\mathcal{O} (10^{12}) \, M_{\odot}$ and each particle has mass $m_p = 10^4 \, M_{\odot}$, so that the Dragon halos are made of $\mathcal{O}(10^8)$ particles.

Fig. \ref{fig:mock_halos_surface_density} shows the surface density map $\rho_s = M_{\rm tot}/A_{\rm pix}$ of Dragon halos representatives of each density profile, $A_{\rm pix}$ being the area of each pixel and $M_{\rm tot}$ the total mass in the corresponding column volume perpendicular to the projection plane. We used the Dragon catalogue to validate our results for the purposes of the present study. In particular, we checked that:

\begin{enumerate}
    \item[i)] {\tt NAJADS} can accurately recover the average radial density distribution for each of our spherical mock halos;
    \item[ii)] the average density is well recovered even in the presence of sub-halos - at least for the most massive ones, which are the most likely to produce DM annihilation signals;
    \item[iii)] the J-factors calculated through {\tt NAJADS} on our mock halos match theoretical expectations.
\end{enumerate}
In the following, we report the results obtained for the first halo of the Dragon catalogue (that we name Dr-1), which follows an Einasto density profile. We always check the consistency of our results across all of the mock halos, and use Dr-1 as an exemplary case. Results for different halos are not reported in the text, unless a relevant difference occurs with respect to Dr-1.

\subsection{Test of the density}
\label{subsec:density_tests}

The first of our validation tests is performed on smooth halos only, i.e.~on mock halos purged of all sub-structures.
We start by randomly drawing $10^3$  LOS within each halo. By LOS we specifically mean a radial line with its origin at the observer position inside the halo, and pointing outwards in a random angular direction.
In this phase of testing, the observer was placed in the halo centre to take full advantage of the symmetries of the matter distribution. 

\begin{figure}
    \centering
    \includegraphics[width=0.95\textwidth]{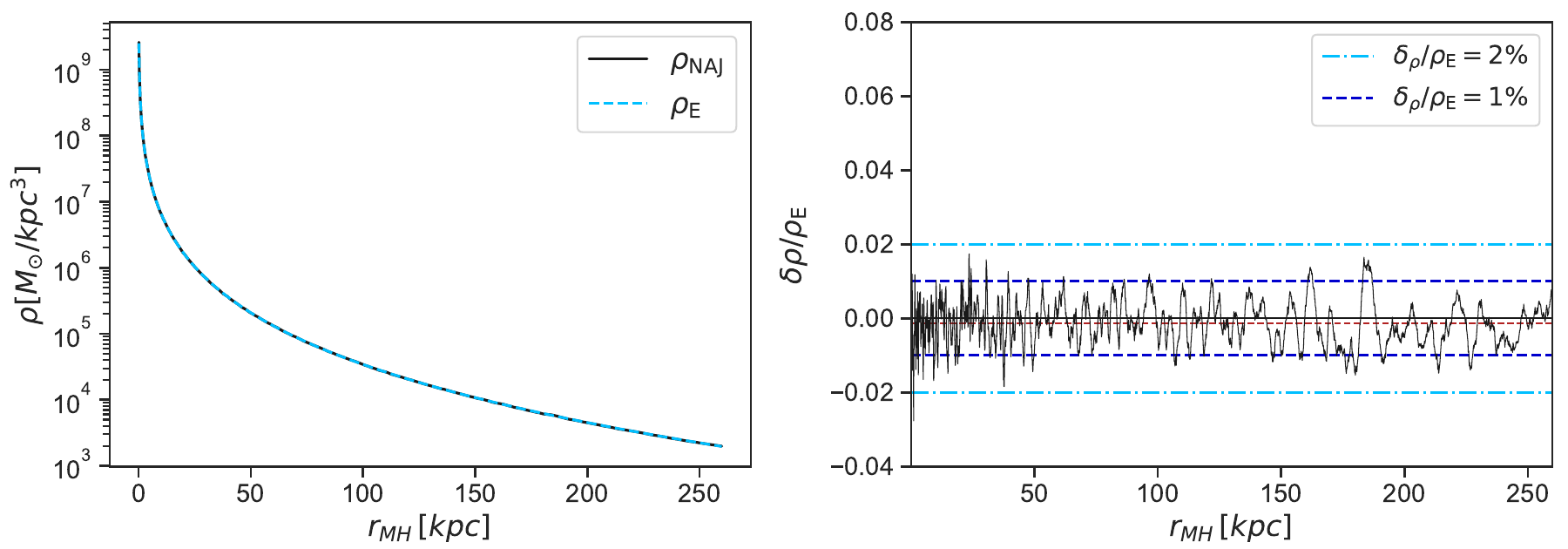}
    \caption{Left panel: comparison between the numerical density profile averaged over $10^3$ LOS (black line) and the corresponding analytical profile (cyan line), for the halo Dr-1 in our MC mock catalogue. Right panel: relative difference $(\rho_{\rm NAJ}-\rho_{\rm E})/\rho_{\rm E}$ between the averaged numerical profile and the semi-analytical one. The blue dashed and cyan dash-dotted lines mark, respectively, the values $0.01$ and $0.02$.}
    \label{fig:profile_mh_E}
\end{figure}

In practice then, drawing $10^3$ LOS simply means extracting $10^3$ couples of angular coordinates $(\theta, \phi)$. Along each LOS, we use {\tt NAJADS} to compute the DM density at equispaced points from $r_{\rm MH} = 0$ to $260$ kpc, with steps of $0.1$ kpc, $r_{\rm MH}$ being the distance from the observer located in the centre. For each LOS we thus obtain a numerical density profile $\rho(r)$ that we average over all LOS. We show our results in the left panel of Fig.~\ref{fig:profile_mh_E}, where we plot the comparison between the averaged numerical profile $\rho_{\rm NAJ}$ (in black) and the analytical one $\rho_{\rm E}$ (in light blue) for the halo Dr-1.
As we work with spherically symmetric halos,  we expect to recover, within some errors, the theoretical density profile of Eq.~\eqref{eq:analytical_profile_a}. We verify that this is indeed the case, and that the correct theoretical density profile for all halos is recovered within $1.5\%$ precision at all radii, as it can be appreciated in the right panel of Fig.~\ref{fig:profile_mh_E}. Here, we plot the relative difference $(\rho_{\rm NAJ}-\rho_{\rm E})/\rho_{\rm E}$, while the blue and cyan line respectively mark the values $0.01$ and $0.02$.\\

Because of spherical symmetry, in principle we would also expect $\rho(r)$ to be constant on all LOS at fixed $r_{\rm MH}$. However, when generating a MC mock halo, local fluctuations of the density can arise, due to the random scatter of the particles that are drawn inside the halo (see App.~\ref{app:mock_catalogue} for more details). This is the reason why we prefer to use the averaged density profile to characterize {\tt NAJADS}, as that is the only quantity that should match, at all radii, the theoretical profile. Indeed, the main consequence of the MC density fluctuations is that $\rho(r_{\rm MH})$ is not the same in all directions. We gauge the entity  of these local fluctuations in the first part of App.~\ref{app:density_scatter}. \\

We repeat the previous steps for sub-halos, and randomly draw again $10^3$ LOS in every smooth halo, this time placing on each LOS the MC realisation of a single sub-halo with fixed mass and the same density profile of its host, but re-simulated for each LOS. The centre of the sub-halo is set on the LOS at $r_{\rm MH} = 120$ kpc\footnote{We apply this procedure to sub-halos in the mass range $10^9-10^7 \, M_{\odot}$. We restrict this test to the higher masses as, given the sensitivity of current detectors, only these are expected to produce an annihilation signal bright enough to be actually seen as resolved gamma-ray sources~\citep{Calore:2019lks,Grand:2020bhk}. We still checked that smaller sub-halos are well recovered by {\tt NAJADS} as local peaks in the density field, down to structures containing a few tens of particles, that become indistinguishable from the typical fluctuations of the main halo.
}.
As we discuss in App.~\ref{app:mock_catalogue}, the MC realisation of a sub-structure is a delicate procedure: especially when modelling the centres of the smallest halos, with a low number of particles, little variations in the number of simulated particles falling in those regions can produce very different outcomes. The purpose of re-simulating the sub-halo for each LOS is to account for this halo-to-halo scatter in our tests. Once again, we run {\tt NAJADS} to compute the DM density over all the LOS for each Dragon halo.

\begin{figure}
	\includegraphics[width=0.325\textwidth]{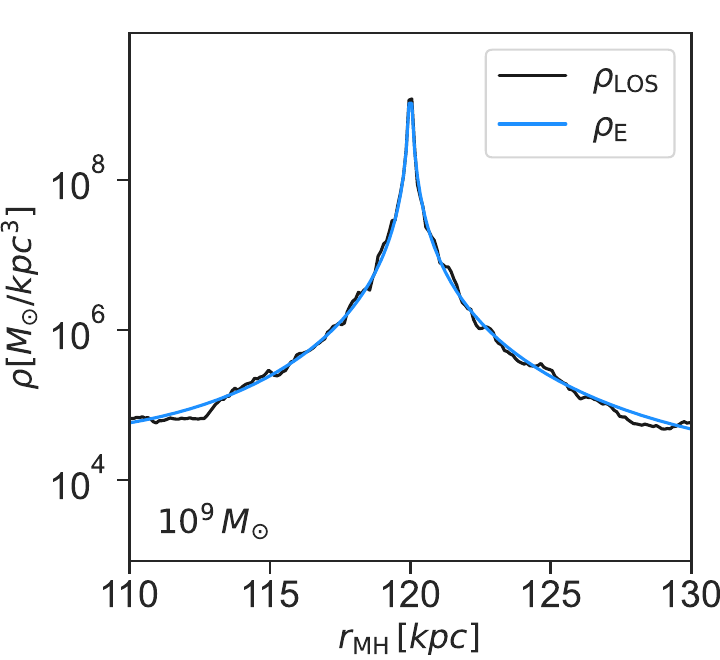}
	\includegraphics[width=0.325\textwidth]{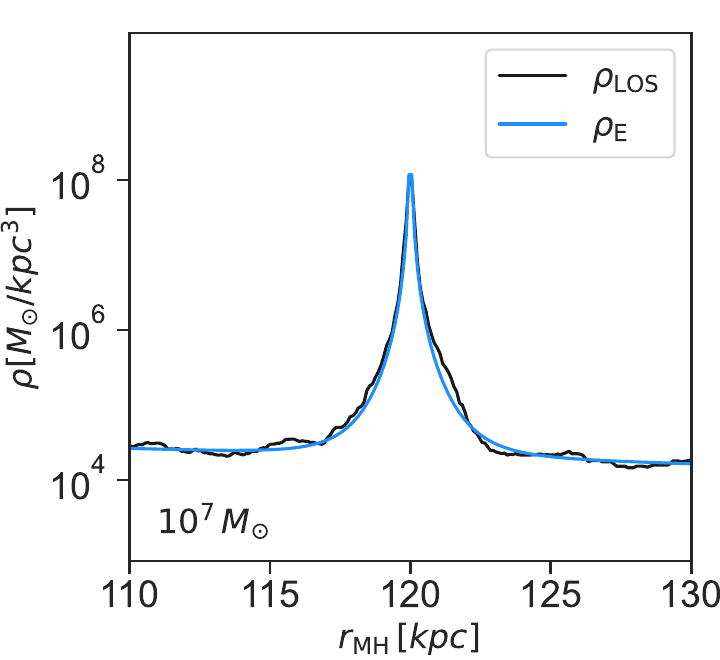}
	\includegraphics[width=0.325\textwidth]{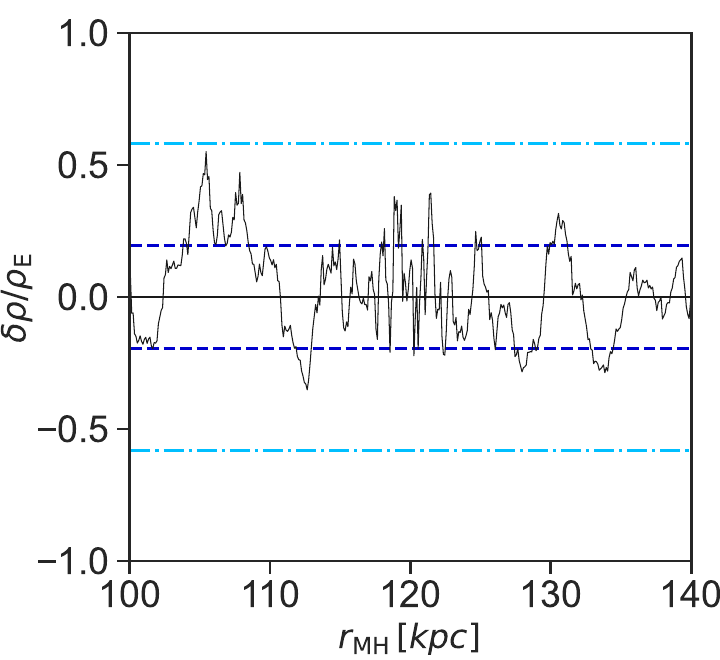}
	\caption{The left panel displays the LOS density profile of the Dr-1 smooth halo, embedding a sub-halo of mass $10^9 \, M_{\odot}$  centred at $r_{\rm MH} = 120$ kpc, zoomed-in on the sub-structure position. For the same sub-halo, the right panel shows the relative difference $(\rho_{\rm LOS}-\rho_{\rm th})/\rho_{\rm th}$, and the $1\sigma$ (blue dashed) and $3\sigma$ (cyan dot-dashed) confidence regions.
    The middle panel is the same as the left panel, but for a sub-halo of mass $10^7 \, M_{\odot}$. }
	\label{fig:profile_subhalos_E}
\end{figure}
The left and central panels of Fig.~\ref{fig:profile_subhalos_E} show the density profiles along a single LOS within the Dr-1 halo, zooming in on the location of a sub-halo of mass $10^9 \, M_{\odot}$ and $10^7 \, M_{\odot}$ respectively. The black line represents the density as computed by {\tt NAJADS} ($\rho_{\rm LOS}$), compared to the theoretical profile $\rho_{\rm E}$ (light blue line). {\tt NAJADS} performance on high-mass sub-structures strikes as remarkably good: we find that for all LOS piercing sub-halos within the mass range $10^9-10^7 \, M_{\odot}$ the density in the outskirts and the central regions of the sub-structure is recovered efficiently. This mass range corresponds, at our resolution, to sub-halos containing 1000 particles or more. The relative discrepancies $(\rho_{\rm LOS} - \rho_{\rm E})/\rho_{\rm E}$ between {\tt NAJADS} computation and the theoretical profile are reported in the right panel of Fig.~\ref{fig:profile_subhalos_E}. The blue and cyan dotted lines mark respectively the $1\sigma$ and $3\sigma$ confidence regions, where $\sigma$ is the density scatter around the analytical profiles that we measure for the main halo (see also Fig.~\ref{fig:sigma_halo+subhalos_E}). We can appreciate that the fluctuations within the sub-structure remain of the same order as those already observed in the main halo, thus likely sharing the same origin of local fluctuations due to the MC procedure. 

When analysing Fig.~\ref{fig:profile_subhalos_E}, we imposed a lower radial cut-off $r_{\rm sh} \geq 0.05$ kpc. 
Plotting the same profiles without cutoff shows that, while the Burkert (cored) profile is still overall well recovered by {\tt NAJADS}, the cuspy analytical profiles (Einasto and NFW) differ significantly from their numerical counterpart in the inner $50$ pc, steeply reaching a peak three orders of magnitude higher.
The limited resolution explains the observed mismatch. Indeed, it is impossible to reproduce a continuous density distribution with a finite number of particles down to arbitrary small radii.
Conceptually, this has important consequences. Using the profiles of Eq.~\eqref{eq:analytical_profile} to study sub-structures requires making an assumption on the density behaviour in the innermost regions of the halo, and then extrapolating such profiles down to $r_{\rm sh} = 0$. On the other hand though, most of these profiles are so steep to span 3 orders of magnitude in density \textit{inside} the minimum convergence radius of the halo. The cusps are not observable in cosmological simulations for most sub-halos, even if it does not necessarily mean that they are not realistic ansatz.
Our approach to J-factor calculation plays a very conservative role in this sense, estimating the density only based on the particles that are actually present in the halos and not relying on additional assumptions.\\

All above tests were repeated on realizations of the halos Dr-1, Dr-3 and Dr-7 at lower ($m_p = 10^5 \, M_{\odot}$) and higher ($m_p=10^3 \, M_{\odot}$) resolution. We remark that the results are unchanged for the main halo, and for sub-halos with the same number of particles as those discussed here. This also confirms the reasonable intuition that the more resolved the simulation, the smaller the DM features that can be resolved with {\tt NAJADS}.

\subsection{Test on the J-factor}
\label{subsec:jfactor_tests}

To show the potentialities of {\tt NAJADS} application to DM indirect probes, we use our catalogue of mock halos to test the J-factor computation.
We compare the J-factors computed running {\tt NAJADS} with the theoretical ones -- obtained by integrating the analytical profiles of Eqs.~\eqref{eq:analytical_profile}. For clarity, we will indicate the theoretical quantities with the subscript ``$\rm th$'' (e.g. $\rho_{\rm th}, J_{\rm th}$). $\rho$ and $J$ will instead be quantities computed through {\tt NAJADS}, unless explicitly specified otherwise. Starting from here and for the rest of the paper we place the observer at a distance $r_{\rm MH} = 8$ kpc from the main halo centre, and refer all radial distances $r$ to this observer.
The difference between using $\rho$ and $\rho_{\rm th}$ for the J-factor is subtle, but substantial. When computing $J$ through {\tt NAJADS}, the density that enters Eq.~\eqref{eq:jfactor_definition} is a local density, that  in general varies if one considers the same LOS in a different realisation of the same halo, due to local perturbations. The analytical density that gives $J_{\rm th}$ instead is an averaged density, so that the equality $\rho_{\rm th} = \langle \rho \rangle$ should be valid, where all average symbols hereafter are intended over different realisations of the same LOS.

One might be tempted to assume that the averaged $J$ should also match $J_{\rm th}$, and in this perspective it becomes interesting to compute the difference between the two. We can write:

    \begin{equation}\label{eq:jfactor_offset}
    \begin{split}
        \frac{\delta J}{J_{\rm th}} &= \frac{\langle J \rangle - J_{\rm th}}{J_{\rm th}} = \frac{\langle \int_{\rm LOS} \rho^2(r, \psi)dr \rangle - \int_{\rm LOS} \rho^2 _{\rm th}(r, \psi) dr}{\int_{\rm LOS} \rho^2_{\rm th}(r, \psi) dr} = \\
         &= \frac{\int_{\rm LOS}(\langle \rho^2 (r, \psi) \rangle - \langle \rho(r, \psi) \rangle^2) dr}{\int_{\rm LOS} \rho_{\rm th}^2(r, \psi) dr} =\frac{\int_{\rm LOS} \sigma_{\rho}^2(r, \psi) dr}{\int_{\rm LOS} \rho_{\rm th}^2(r, \psi) dr} \, ,	
    \end{split}
    \end{equation}
from which one can deduce that there is an appreciable difference to be expected between $\langle J \rangle$ and $J_{\rm th}$. In computing $J$, we keep track of local fluctuations in the density, and because the average of the squared density is higher than the square of the average density, the numerical averaged J-factor is higher than the theoretical value. 
This means that using a smooth analytical profile is substantially different from evaluating the peculiar density point-by-point along the LOS. A related phenomenon is already well known in the literature as boosting~\citep{Kuhlen:2012ft}. Boosting occurs typically when an unresolved population of sub-structures is present in the halo. This unresolved population increases the graininess of the halo -- and hence, the density dispersion $\sigma_{\rho}$ -- and can lead to an enhancement of the ``true'' J-factor with respect to $J_{\rm th}$. 

We detail in App.~\ref{app:density_scatter} that, for our mock smooth halos, the relative dispersion $\sigma_{\rho}/\rho_{\rm th}$ remains approximately the same in all regions of the halo. We can then push the computation of Eq.~\eqref{eq:jfactor_offset} a little further and write: 

    \begin{equation}
        \label{eq:jfactor_offset_MC}
         \frac{\delta J}{J_{\rm th}} =\frac{\int_{\rm LOS} \sigma_{\rho}^2(r, \psi) dr}{\int_{\rm LOS} \rho_{\rm th}^2(r, \psi) dr} = \frac{\int_{\rm LOS} (\sigma_{\rho}^2 /\rho_{\rm th}^2) \rho_{\rm th}^2(r, \psi) dr}{\int_{\rm LOS} \rho_{\rm th}^2(r, \psi) dr} = \frac{\sigma_{\rho}^2}{\rho_{\rm th}^2} \, .
    \end{equation}

As we find $\sigma_{\rho} / \rho_{\rm th} \simeq 0.194$, the J-factor offset can be estimated as $\delta J/J_{\rm th} \simeq (0.194)^2 \sim 3.76 \%$. 
With this theoretical prediction at hand, we choose a random LOS in each smooth halo in the Dragon catalogue, and compute the J-factor offset for $10^3$ different realisations of such LOS. In practice, we exploit the spherical symmetry of our halos: we observe that to all LOS at angular distance $\psi$ from the halo centre corresponds the same density profile $\rho_{\rm th}(r, \psi)$. They can then be considered as different realisations of the same LOS, since the numerical density along each LOS will in general fluctuate.

\begin{figure}
	\centering
    \includegraphics[width=0.318\textwidth]{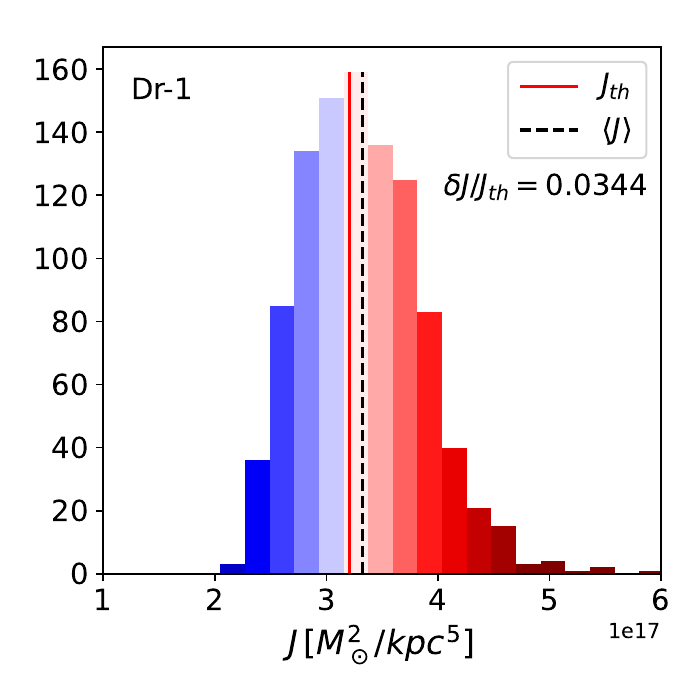}
	\includegraphics[width=0.318\textwidth]{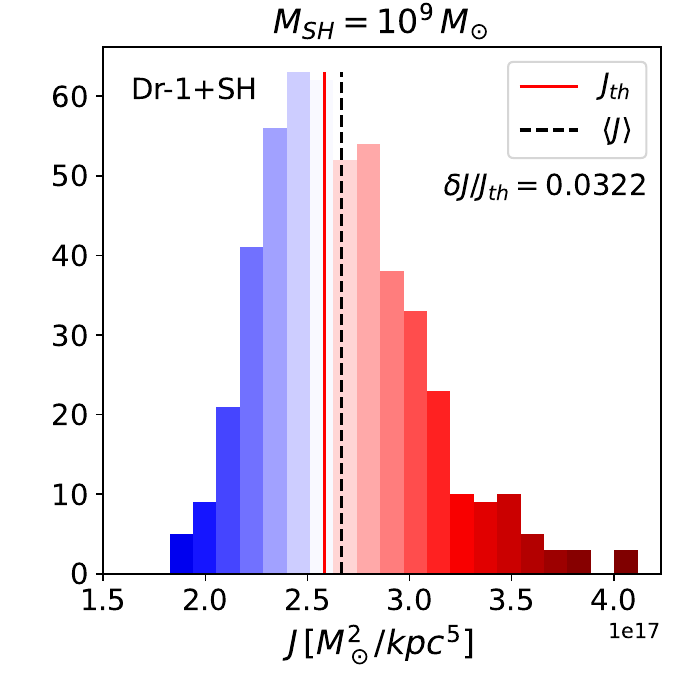}
	\includegraphics[width=0.346\textwidth]{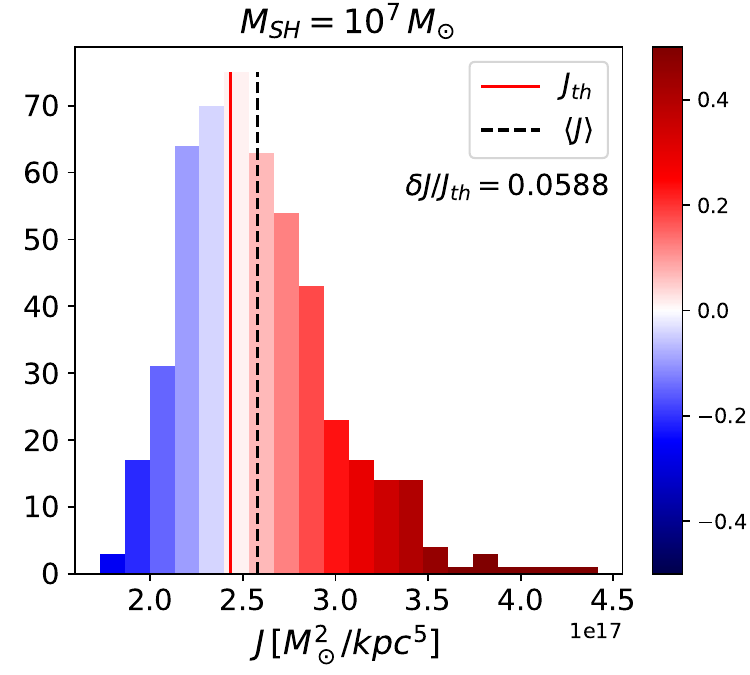}
     \includegraphics[width=0.318\textwidth]{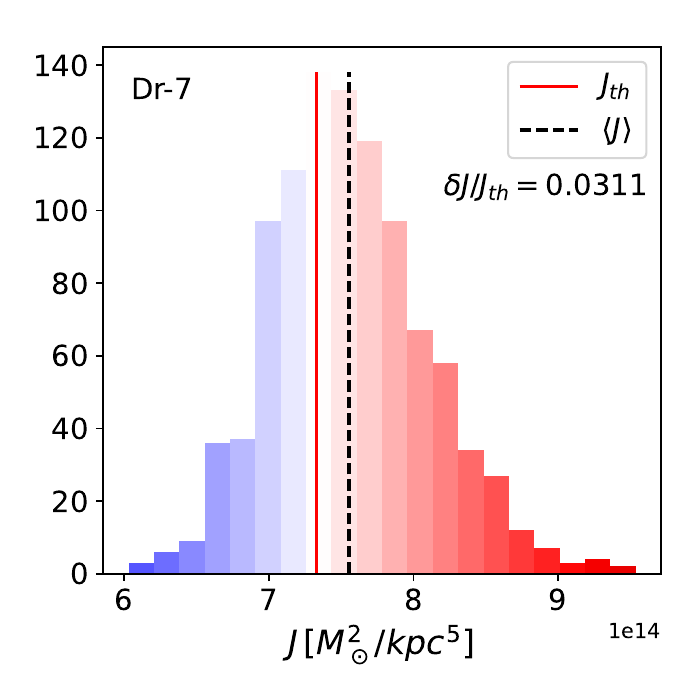}
	\includegraphics[width=0.318\textwidth]{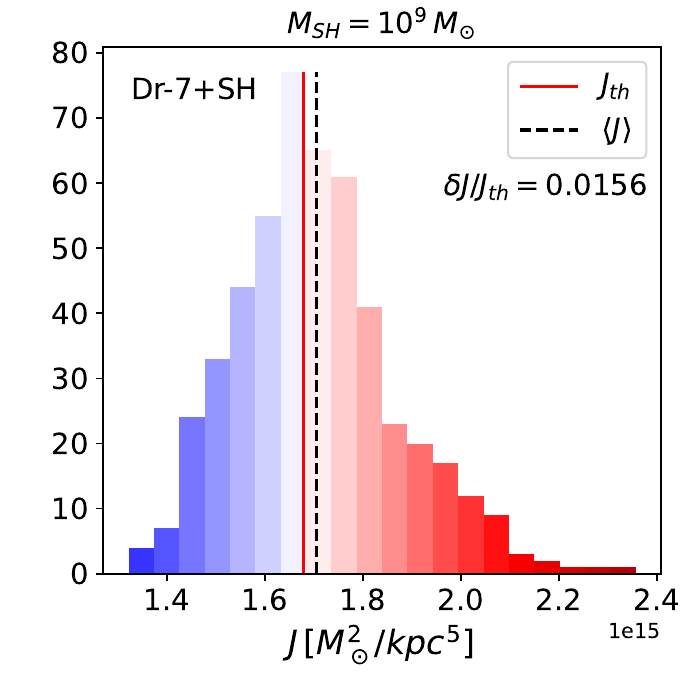}
	\includegraphics[width=0.346\textwidth]{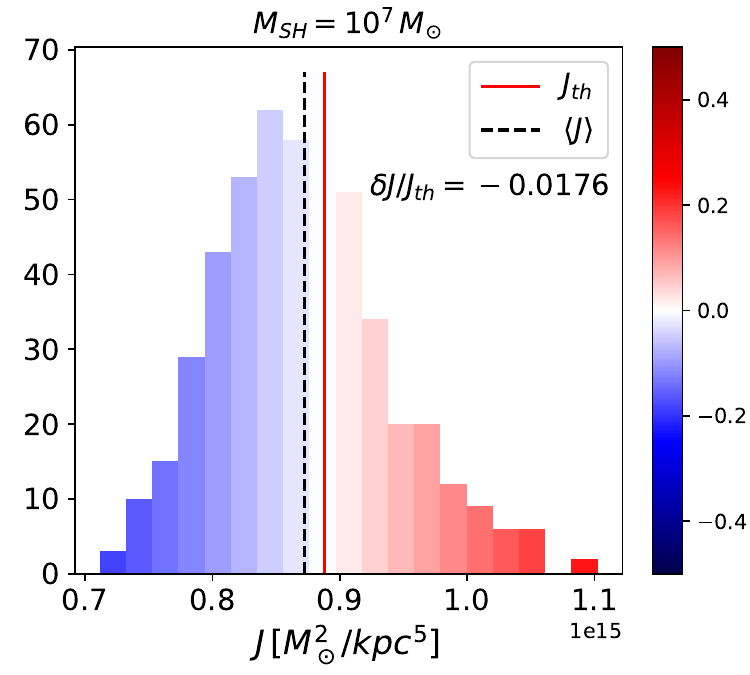}

	\caption{Top left panel: J-factor histogram realized computing J-factors with {\tt NAJADS} along $10^3$ randomly oriented LOS in the smooth halo of Dr-1. Middle and right panel: same as left panel, but with each LOS piercing a different MC realisation of a sub-halo of mass $10^9 \, M_{\odot}$ and $10^7 \, M_{\odot}$ respectively. The black dashed line marks the histogram average, the red solid line marks the expected theoretical value. The colour scale indicates the $\%$ difference between the value of each bin $J_{\rm bin}$ and the average of the distribution. The lower row is the same but for the  Burkert profile (Dr-7).}
	\label{fig:Jfactor_histograms_E}
\end{figure}

In the first panels of Fig. \ref{fig:Jfactor_histograms_E} are shown the histograms of the $10^3$ J-factor values obtained for the LOS drawn in the smooth halo Dr-1 (top row) and Dr-7 (bottom row). The colour scale marks the single bin offset $(J_{\rm bin}-J_{\rm th})/J_{\rm th}$, while the mean value $\langle J \rangle$ and $J_{\rm th}$ are marked in the plots with vertical dashed black and solid red lines respectively. The average offset value is indicated as $\delta J/J_{\rm th}$ and amounts to $3.44 \%$ ($3.11 \%$), in reasonable agreement with the predicted $3.76\%$. 
Though a difference between $\langle J \rangle$ and $J_{\rm th}$ of $\sim 3-4 \%$ is a negligible systematic error compared to the speed and simplicity of modelling DM halos with analytical profiles, we should recall an essential consideration: the constant value of $\sigma_{\rho}/\rho \sim 0.194$ does not hold the same for halos in cosmological simulations. There are numerous physical effects contributing to the graininess of simulated halos that, depending on their formation history, can easily enhance the density scatter up to much higher values (as we further discuss in App.~\ref{app:density_scatter}). Indeed, the boosting of the J-factor signal occurring in real halos has been  estimated to reach up to $200-300\%$ ~\citep{Bartels:2015uba, Grand:2020bhk}.
If DM signature studies rely solely on $J_{\rm th}$, all boosting effects need to be modelled and added independently after the calculation (see e.g.~\cite{Ando:2019xlm} for a review). In this context, exploiting an approach similar to the one we propose here allows to automatically account for boosting without requiring separate modelling.
The histograms in the second and third panels (both rows) of Fig.~\ref{fig:Jfactor_histograms_E} are similar to ones in the left panels, though now each LOS pierces the realisation of a sub-halo of mass $10^9 M_{\odot}$ and $10^7 \, M_{\odot}$, and same density profile of the parent halo.
We still notice the offset $\delta J$ between the average J-factor and the theoretical value, however, this offset is now further affected by the presence of the sub-structure, depending on its mass and profile.
	 
Here, we can fully appreciate the effects of {\tt NAJADS} providing a conservative estimate of the sub-halos central densities. This underestimate counteracts the signal boosting we described earlier: its effect is to reduce $\langle J \rangle$ and hence $\delta J$. However, for halos in the high mass range ($10^8 - 10^9 \, M_{\odot}$) its impact is very small: by comparing the values of $\delta J$ with and without sub-structures we infer that when introducing high-mass sub-halos along the LOS $\delta J$ is reduced of only $\sim 0.2\%$ with respect to the smooth case, for cuspy profiles (both Einasto and NFW). Cored profiles are more affected due to the lower number of particles that can be found in the centre of sub-structures (see App.~\ref{app:mock_catalogue}). Indeed, for Burkert halos, we observe a $\delta J$ reduction of$\sim 1.5\%$.
This is overall consistent with the consideration of Sec.~\ref{subsec:density_tests} that {\tt NAJADS} density estimates on high-mass structures are robust and less affected by the MC Poissonian noise, even in the inner regions.
As shown in App.~\ref{app:density_scatter} (see  Fig.~\ref{fig:sigma_halo+subhalos_E}) for sub-halos of masses $10^7 \, M_{\odot}$ the density scatter $\sigma_{\rho}/\rho$ at the halo centres increases from $\sim 20 \%$ up to $60 \%$. Even if variations in the density scatter are confined to a very small portion of the LOS, the numerical J-factor computed for these sources will be affected, as a consequences of Eq~\eqref{eq:jfactor_offset}. This is what is happening in the right panel of Fig. \ref{fig:Jfactor_histograms_E}, where for sub-halos with cuspy profiles, $\delta J$ is $\sim 2.4 \%$ higher when the LOS pierces a sub-halo.

All J-factors presented in Fig.~\ref{fig:Jfactor_histograms_E} were computed with a radial cutoff $r_{\rm sh} \geq 0.05$ kpc. When the cut-off is removed and contributions from $\rho(r_{\rm sh} \rightarrow 0)$ are included in the picture, the results change drastically. $\delta J$ now becomes negative (i.e. the analytical J-factor is always much higher than the averaged numerical one) for every combination of halo + sub-halo, as the steep theoretical profiles predict a signal a good $10 \%$ higher than the numerical J.

Trusting or not that DM in sub-halos would actually form steep cusps if we had enough particles available to resolve them, it looks undeniable that relying solely on analytical profiles to compute annihilation fluxes can have substantial effects.

\section{The proof of concept: J-factor skymaps}
\label{sec:proof_of_concept}

In this Section, we will show two practical end-to-end applications of {\tt NAJADS}.
Specifically, we use it to generate HEALPix \citep{Gorski:2004by} celestial maps of the J-factor for the halos in the Dragon catalogue and for one halo from a cosmological simulation.
We compare these maps with the corresponding semi-analytical ones. As a double check, all semi-analytical maps in this section were also separately computed with a dedicated tool, {\tt CLUMPY}~\citep{Charbonnier:2012gf,Bonnivard:2015pia,Hutten:2018aix}.

In what follows, we refer separately to {\it smooth halos} -- from which sub-halos are subtracted -- and {\it full halos} containing the entire sub-halos population.

\subsection{J-factor calculation for the Dragon catalogue's halos}
\label{subsec:skymaps}

We start with the Dragon catalogue. We treat each halo as a mock version of the Milky Way halo and produce celestial maps of the J-factor as seen from an observer located on the Galactic plane, at a distance of 8 kpc from the halo centre. Each map has ${\rm NSIDE}=128$, corresponding to 196608 pixels.

\begin{figure}
    \includegraphics[width=0.5\textwidth]{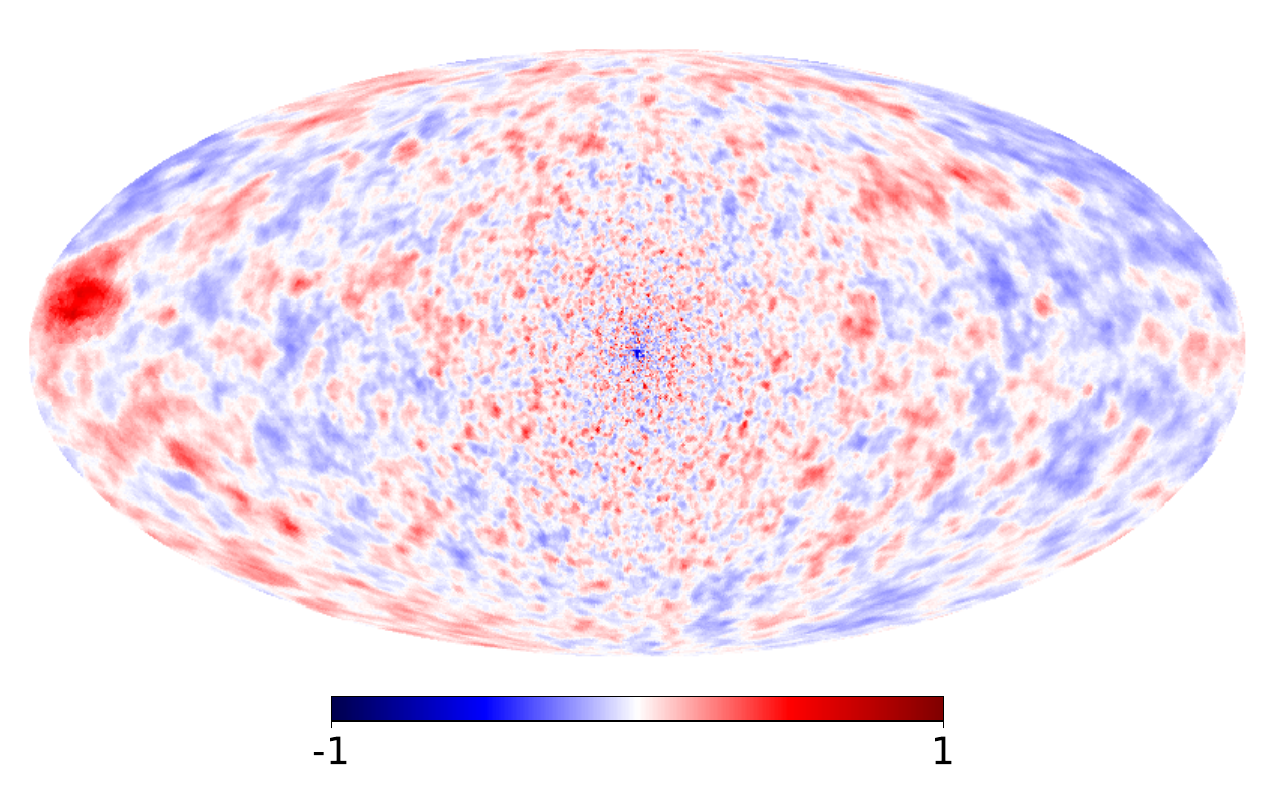}
    \includegraphics[width=0.5\textwidth]{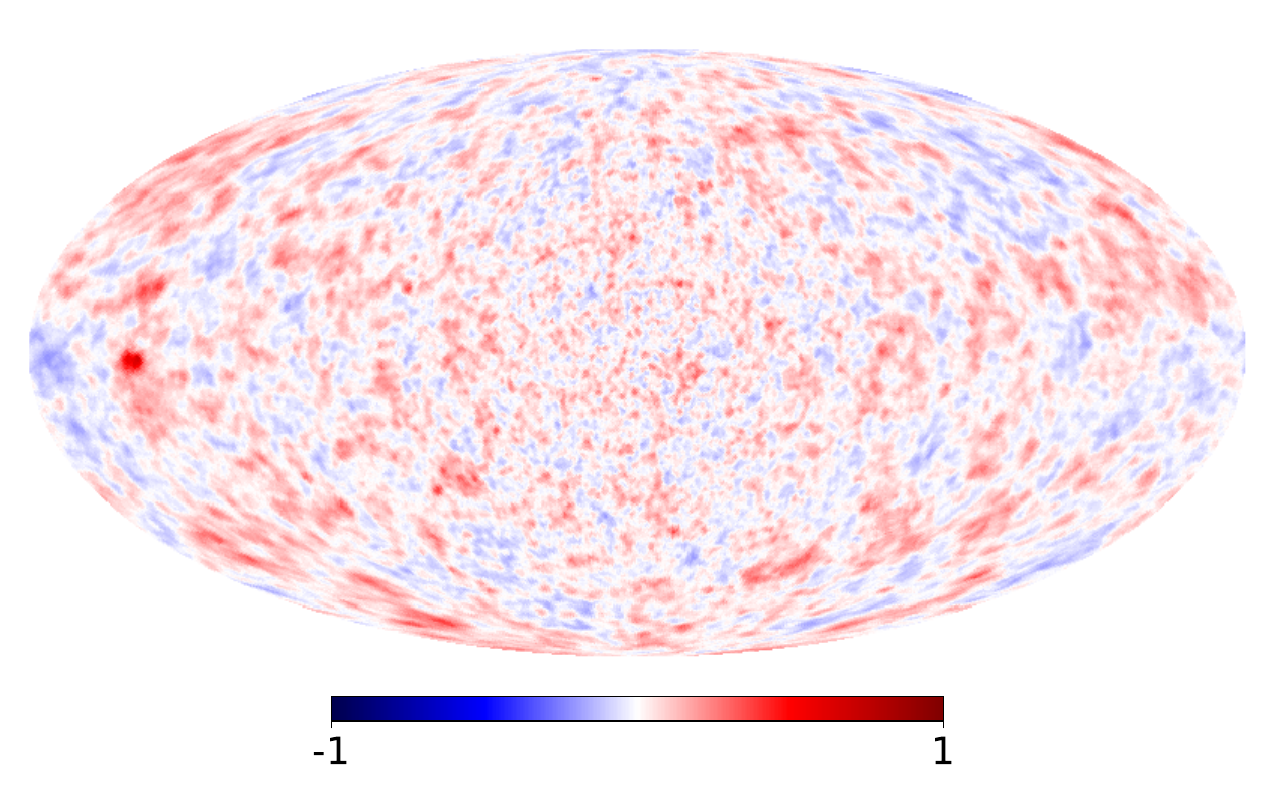}
    \caption{Residual J-factor sky maps for the mock halos Dr-1 (Einasto profile, left panel) and Dr-7 (Burkert profile, right panel), including both the main-halo and the sub-halos contributions. The residual map is obtained as $(J-J_{\rm th})/J_{\rm th}$, where $J_{\rm th}$ is the J-factor map computed integrating the analytical profiles of Eq.~\eqref{eq:analytical_profile} and $J$ is the map computed with {\tt NAJADS}.} 
    \label{fig:residuals_mock}
\end{figure}

Fig. \ref{fig:residuals_mock} shows the residual J-factor sky maps for the Dragon halos Dr-1 (left panel) and Dr-7 (right panel).
These maps refer to full halos, although they are strongly dominated by the main-halo component.
They portray the relative difference between the map obtained with {\tt NAJADS} and the corresponding semi-analytical map.
This residual maps thus provide a visualisation of the DM anisotropies in our mock galactic-sized halos.
Mapping these fluctuations is a distinctive advantage of {\tt NAJADS} that is not available when J-factor maps are produced with traditional techniques. 
We remark that the exact morphology of these fluctuations depends, of course, on the invidual halo realization. However, they present the same statistics as the histograms of Fig.~\ref{fig:Jfactor_histograms_E} as they share the same origin, i.e. the MC noise (see also Sec.~\ref{subsec:jfactor_tests}). As such, they remain overall relatively small, mostly $\leq 25\%$. We will show how this changes when we run {\tt NAJADS} on a more realistic halo.

\subsection{An application to a realistic halo in IllustrisTNG}
\label{subsec:TNG}

For {\tt NAJADS}' final proof of concept, we have selected one halo from the TNG50-1-Dark simulation in the IllustrisTNG\footnote{\url{https://www.tng-project.org/}} suite \citep{Pillepich:2017jle,Nelson:2018uso,Nelson:2019jkf,Pillepich:2019,Pillepich:2019bmb}.
TNG50-1-Dark is a DM-only realisation of a 35 Mpc/$h$ side cube, with a total of about 10 billions DM particles of mass $\sim 3.65 \cdot 10^{5} \, M_{\odot}/h$ each. The cosmological parameters have been fixed according to Planck 2015 data \citep{Planck:2015fie}.  All halos and sub-halos have been identified with the FoF and {\tt SUBFIND} \citep{Davis:1985rj, Springel:2000qu}  halo finders, respectively.\\

For our study we choose halo 154112 (hereafter TNG-F). This is the third most massive halo of the simulation, with about $1.3 \cdot 10^8$ DM particles, with a total mass of $4.68 \cdot 10^{13} M_{\odot}/h$. TNG-F is a cluster-sized halo, with a virial radius\footnote{Defined as the radius enclosing an average density 200 times higher than the Universe's critical density, $\rho_c = 3H_0^2/8\pi G$.} $\sim 830$ kpc. As we want to keep the focus on applications of {\tt NAJADS} to galactic-like halos and sub-halos, we re-scale the chosen TNG halo to make its properties more similar to those of a typical galactic halo. Specifically, we shrink all lengths by a factor of $5$ and all masses by a factor of $125$, so as to keep all densities unchanged. None of our conclusions on {\tt NAJADS}' performance are in any way affected by this manipulation. After the re-scaling, the total halo mass measures $\sim 5.5 \cdot 10^{11} \, M_{\odot}$.

\begin{figure}
    \centering
    \includegraphics[width=\textwidth]{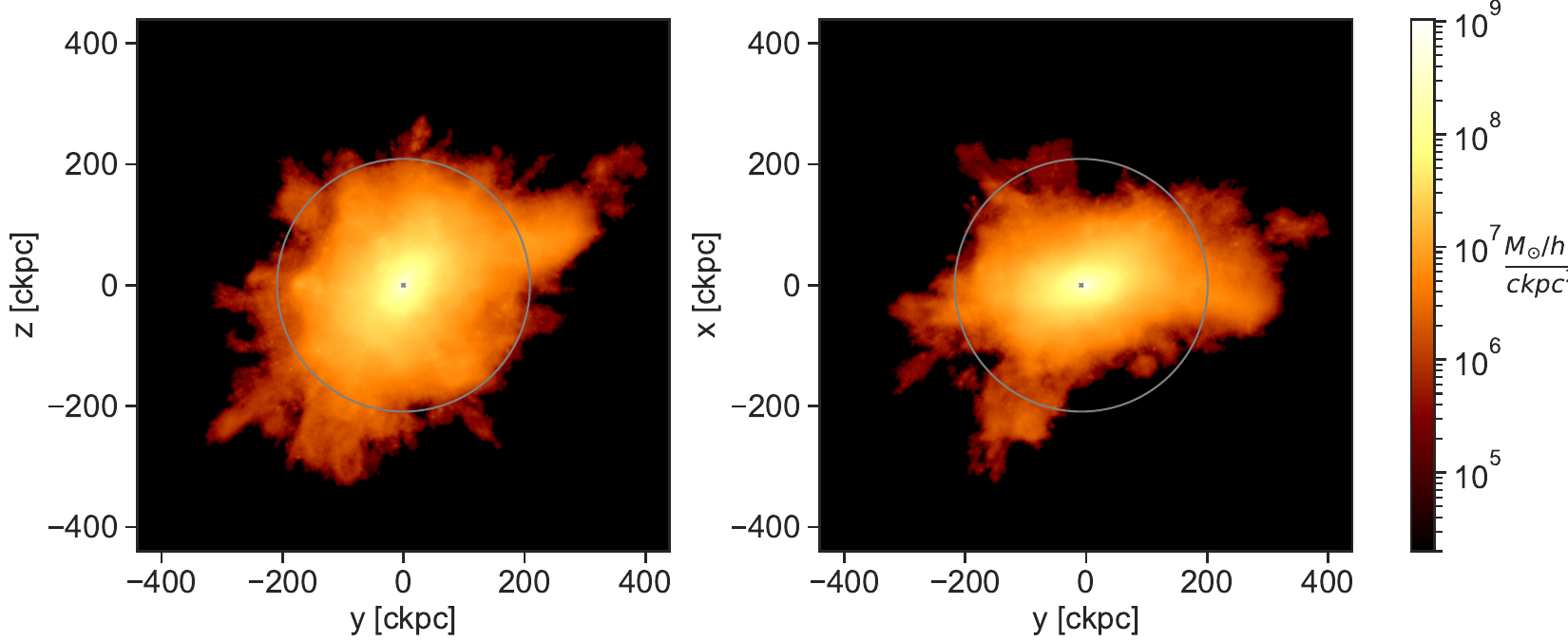}
    \caption{Surface density maps of the halo TNG-S (smooth component). The left and right panels show projections of the halo along the $x$ and $z$ axis, respectively (in comoving units). The grey crosses show the position of the observer at 8 kpc from the centre of the halo on the $x$ axis, while the grey circles mark a region of radius 2.5 times the virial radius, centred on the halo centre.}
    \label{fig:TNG50_surface_density}
\end{figure}

Fig.~\ref{fig:TNG50_surface_density} shows the maps of the integrated surface density of the $z=0$ snapshot of the simulation, zoomed-in on the re-scaled smooth halo. That is, the halo made up of all particles not attributed to a smaller sub-structure by {\tt FoF} or {\tt SUBFIND}, that we then name TNG-S. We choose the reference frame so that the centre of the halo has coordinates $(0,0,0)$ kpc. Panels left and right of Fig.~\ref{fig:TNG50_surface_density} display the projection of the halo along axes $x$ and $z$, respectively, with the grey circle marking $2.5$ times $r_{1/2} = 83.58$ kpc, defined as the radius enclosing half of the total halo mass.
We select this particular halo because, among the most massive ones in TNG50-1-Dark -- and hence better resolved -- it is also the one whose shape looks closest to spherical. However, as it can be seen clearly in the figure, TNG-S still presents a quite evident prolate shape, and its mass distribution is highly asymmetrical. Though this is more common for cluster-sized halos and asymmetries in galactic halos are typically less prominent, especially when baryons are included, they are nonetheless a feature common to every cosmological simulation. We deem it particularly interesting to show how {\tt NAJADS} behaves in similar conditions.\\

\begin{figure}
    \centering
    \includegraphics[width=0.96\textwidth]{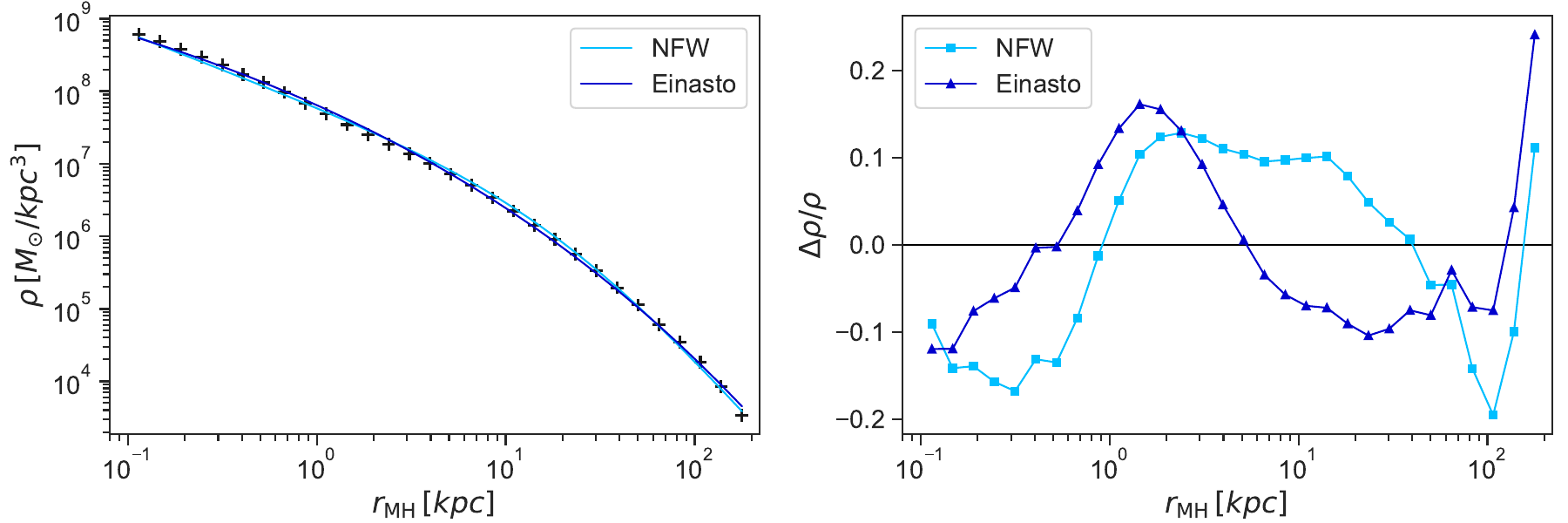}
    \caption{Left panel: Fit of the smooth component density profile of the TNG-S halo, as a function of the distance from its centre. The data were fitted with a NFW profile and an Einasto profile. Right panel: density residuals for the two fits.}
    \label{fig:TNG50_profile_fit}
\end{figure}

We proceed to fit the smooth halo with an analytical mass profile, assuming spherical symmetry to be a good description of the halo, at least to a first approximation. This is oftentimes done in DM phenomenology studies, due to the extreme simplicity of the spherical profiles, which allows for broad and fast analyses. We bin TNG-S in 30 logarithmic radial bins from $r_{\rm MH}=0.1 \, {\rm kpc}$ to $r_{\rm MH} = 210 \, {\rm kpc}$, leaving out the more irregular outskirts. We compute the average density in each spherical shell and fit the resulting data points. Fig.~\ref{fig:TNG50_profile_fit} shows the resulting best-fit curves. We use Einasto and NFW profiles parametrised as in Eqs.~\eqref{eq:analytical_profile}, with three and two free parameters respectively. We find that both curves provide a good fit to the averaged profile, with no strong preference between the two, despite the higher flexibility of the Einasto parametrisation. In all our analytical calculations, we then choose to use for the background halo the Einasto mass profile with parameters $\rho_{-2} = 4.13 \cdot 10^{5} \, M_{\odot}/$kpc$^3$, $r_{-2} = 25.96$ kpc and $\alpha = 0.165 $.\\

\begin{figure}
    \includegraphics[width=0.5\textwidth]{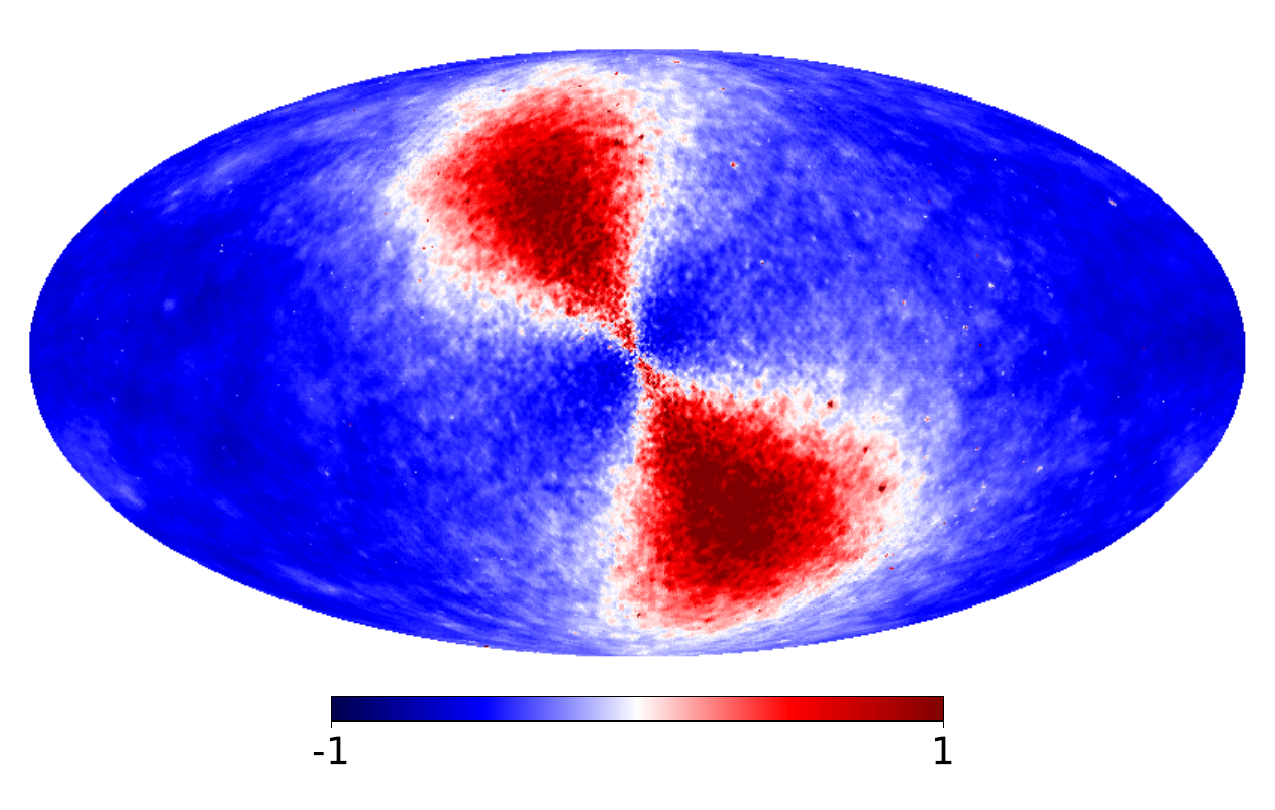}
    \includegraphics[width=0.5\textwidth]{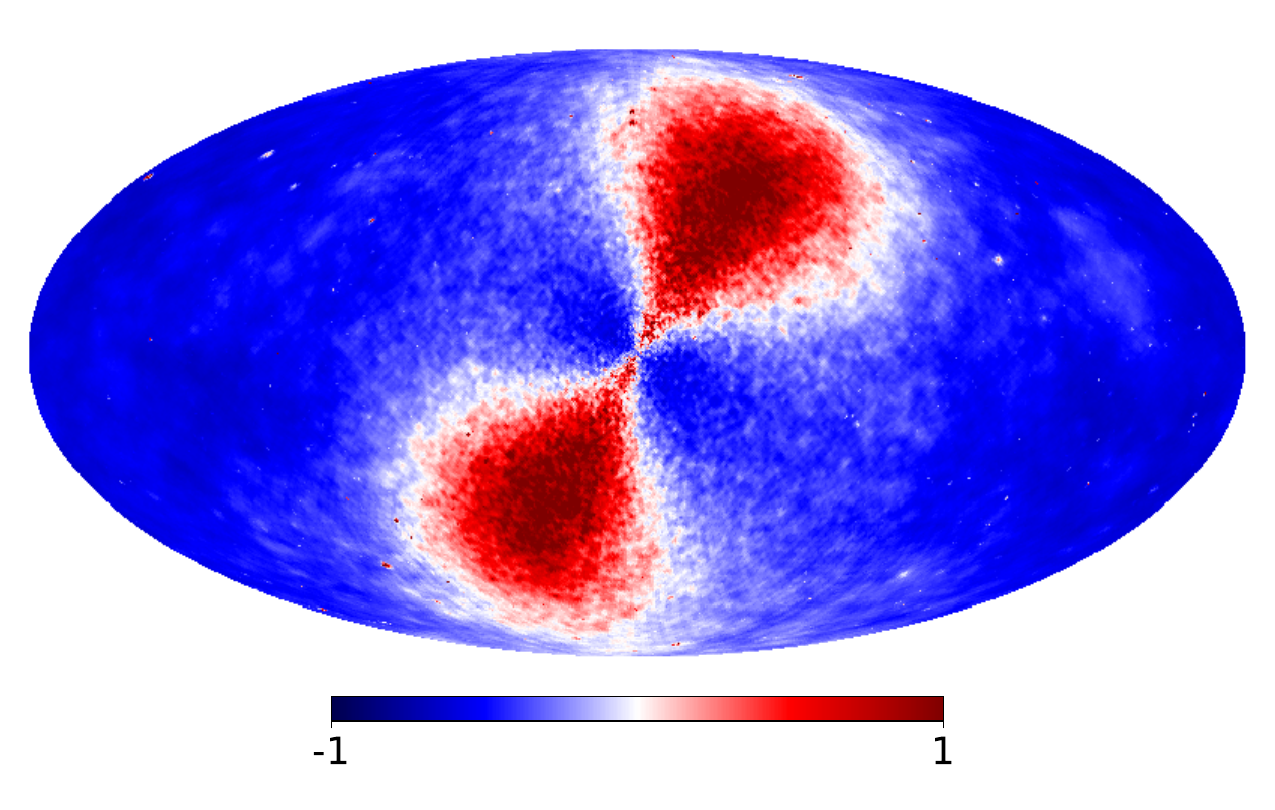}
    \vspace*{0.5cm}
    \includegraphics[width=0.5\textwidth]{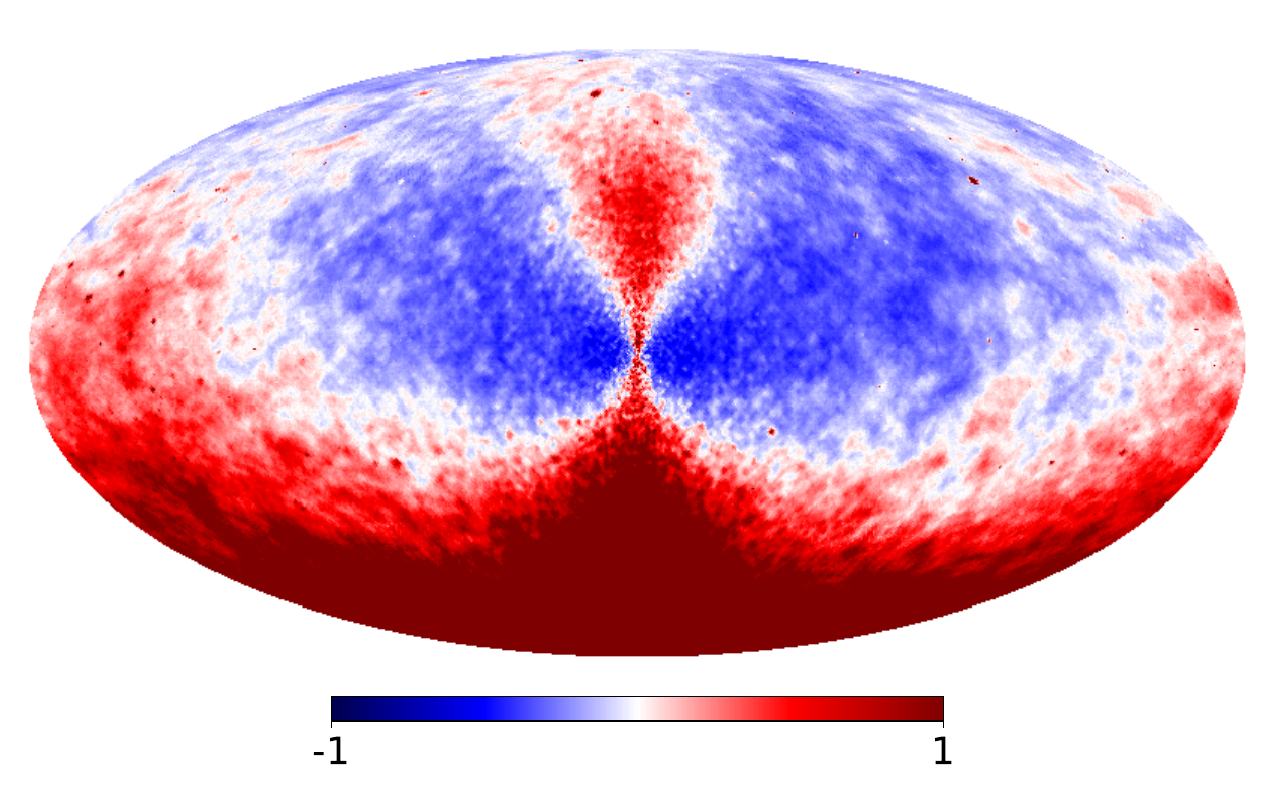}
    \includegraphics[width=0.5\textwidth]{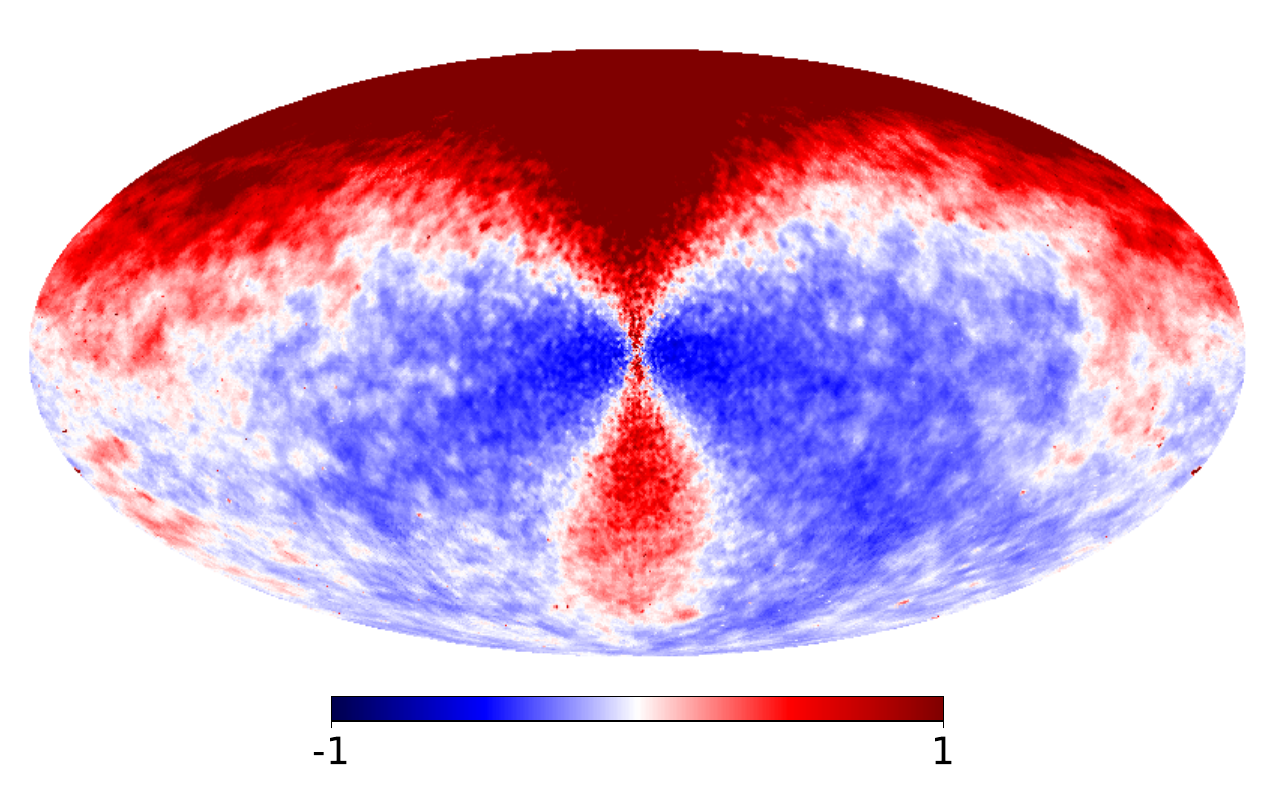}
    \caption{Residual sky-maps of the smooth halo TNG-S computed as $(J -J_{\rm th})/J_{\rm th}$, where $J_{\rm th}$ is the J-factor map obtained through a numerical integration of the best-fitting Einasto profile of Fig. \ref{fig:TNG50_profile_fit}. Different plots were realised placing the observer at different positions in the main halo and re-running {\tt NAJADS} for each configuration: in halo-centric coordinates, the observer was located respectively at $[8.0, 0.0, 0.0]$ kpc, $[-8.0, 0.0, 0.0]$ kpc (top row), $[0.0, 8.0, 0.0]$ kpc, and $[0.0, -8.0, 0.0]$ kpc (bottom row).}
    \label{fig:residuals_TNG50}
\end{figure}

We next run {\tt NAJADS} on the smooth halo TNG-S and compute the J-factor maps, while simultaneously obtaining the theoretical map integrating the analytical background profile up to $r_{\rm MH}=210 \, \rm{kpc}$. The position of the observer is chosen to be $(-8.0., 0.0., 0.0)$ kpc in halo-centric coordinates. To compare the two maps, we calculate the residual map as  $(J -J_{\rm th})/J_{\rm th}$, displayed in the first (top, left) panel of Fig.~\ref{fig:residuals_TNG50}. We can immediately notice a striking difference between this plot and the panels in Fig.~\ref{fig:residuals_mock}, both in terms of intensity and morphology of the fluctuations. Instead of presenting as a statistical white background noise, the residuals have a very distinctive shape, consequence of the fact that we are attempting a fit with a spherical profile on a halo that in reality presents a much more articulate morphology.
What from Fig.~\ref{fig:TNG50_profile_fit} looked like a reasonably good description of the halo, misses out the actual local distribution of DM. On the contrary, {\tt NAJADS} is not blind to these features and is capable of reproducing in detail the shape of the halo. The resulting differences between the numerical J-factors and the semi-analytical ones can reach up to $> 300\%$, depending on the LOS.

The joint analysis of Figs.~\ref{fig:TNG50_surface_density} and~\ref{fig:residuals_TNG50} raises the question whether the introduction of a more complex 3D structure in the analytical fit might ease the tension between analytical and numerical J-factor maps. Though that might, in principle, be the case, we note that introducing a tri-axial structure brings in a much higher level of complication: properly including, for example, a radially-dependent ellipticity term requires first to run a numerical density estimator on the whole halo to identify the shape and orientation of the iso-density curves, after which a complex profile can be iteratively fitted on the halo. This procedure is time-consuming, and rarely adopted in DM phenomenology studies, where this fit would need to be repeated one-by-one over several halos and tens of thousands of sub-halos. On the other hand, once a density estimator has been run on the simulation to determine the iso-density curves, then the fully-numerical approach of NAJADS becomes much more advantageous for the J-factor calculation.

It is important to stress that the specific morphology of the mass anisotropies in the first panel of Fig. \ref{fig:residuals_TNG50} is by no means general. This is better illustrated in the remaining panels of the figure, where we recompute the same J-factor maps moving the observer around in the main halo. Specifically, we perform three more runs with {\tt NAJADS}, placing the observer respectively at locations $(8.0, 0.0, 0.0)$, $(0.0, -8.0, 0.0)$ and $(0.0, 8.0, 0.0)$ kpc. We conveniently keep the observer always at the same distance of 8 kpc from the halo centre, so that the theoretical map $J_{\rm th}$ remains unchanged. However, the residual map transforms significantly. Observers at different locations see the mass anisotropies from different angles, and the morphology of the resulting J-factor map changes accordingly.
This has deep implications: it shows clearly that describing real halos with spherical or, more generally, smooth profiles to predict J-factor signals carries a non-negligible error. This error depends on the mass distribution of the peculiar halo, but can in principle be very high. Indeed, even if smooth profiles are proven to be great \textit{average} descriptions of halos, predictions need to account for the fact that when observing objects in the sky, we do not have the chance of mediating over different points of view. The same error affects the results coming from {\tt NAJADS}: even if the code is sensitive to the local mass distribution, simulated halos are not exact mirror images of the real ones, so in general the specific morphology of mass anisotropies will differ. Whatever the method used to compute J-factors, there is no way around this, unless one is able to gain information on the specific DM morphology of the target halo for which predictions are made.\\
However, we point out that the code that we are developing provides an optimal tool to study the entity and impact of this uncertainty, allowing to quantify, statistically, how often particularly high residuals are recovered in simulated halos with observers at different locations.\\

\begin{figure}
    \includegraphics[width=0.5\textwidth]{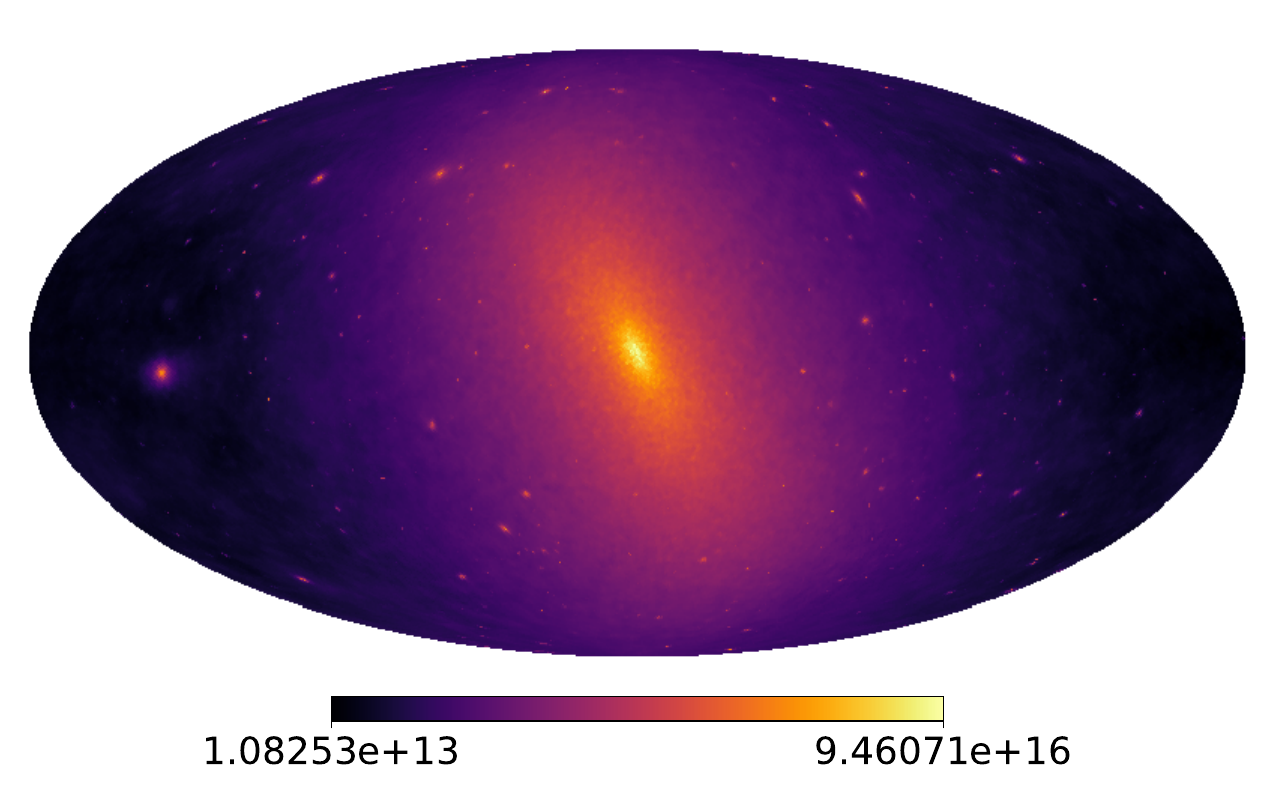}
    \includegraphics[width=0.5\textwidth]{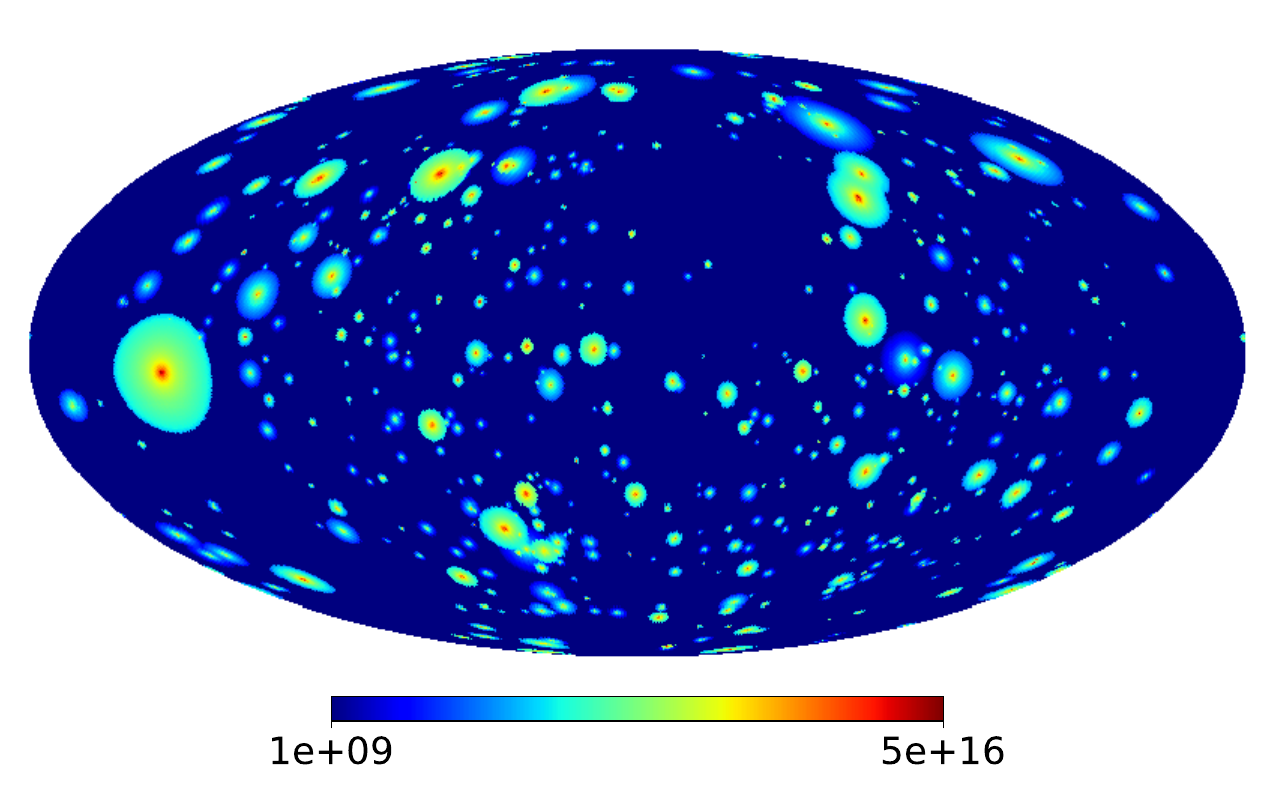}   
    \includegraphics[width=0.5\textwidth]{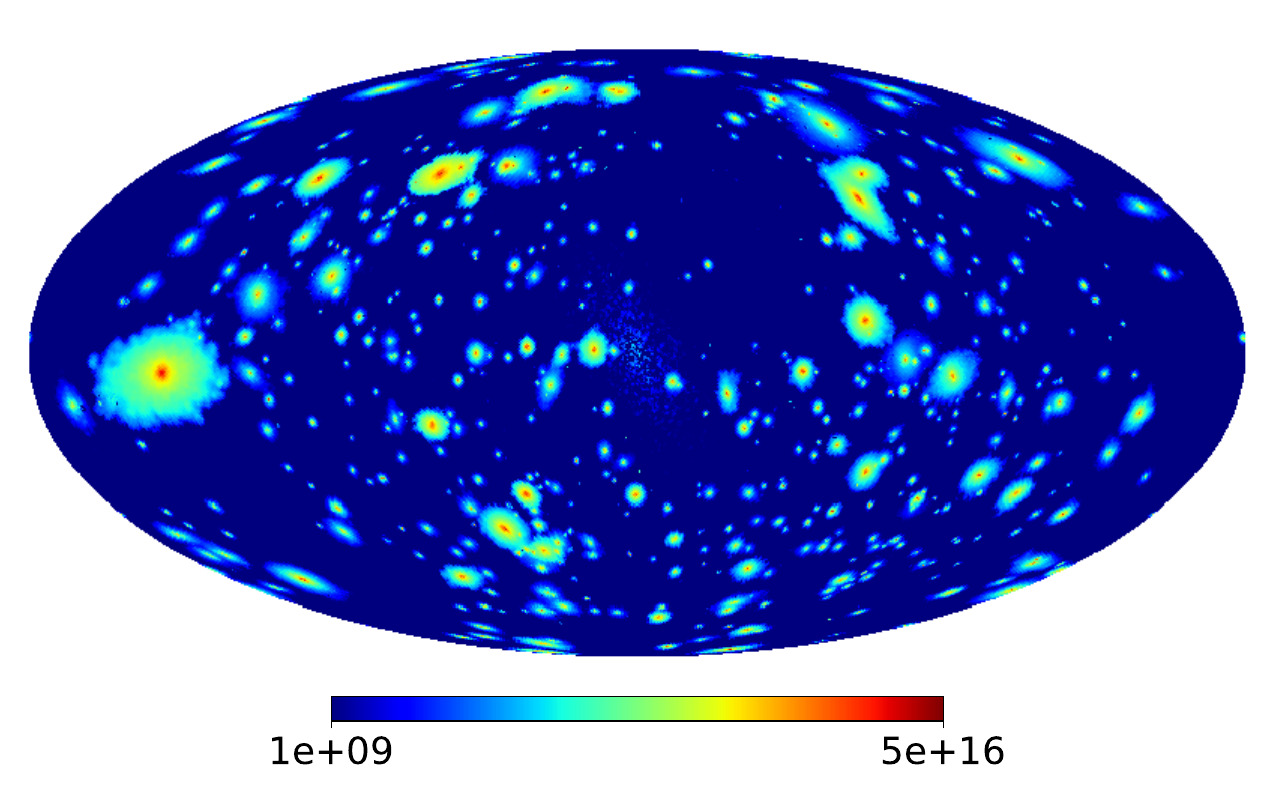}
    \includegraphics[width=0.5\textwidth]{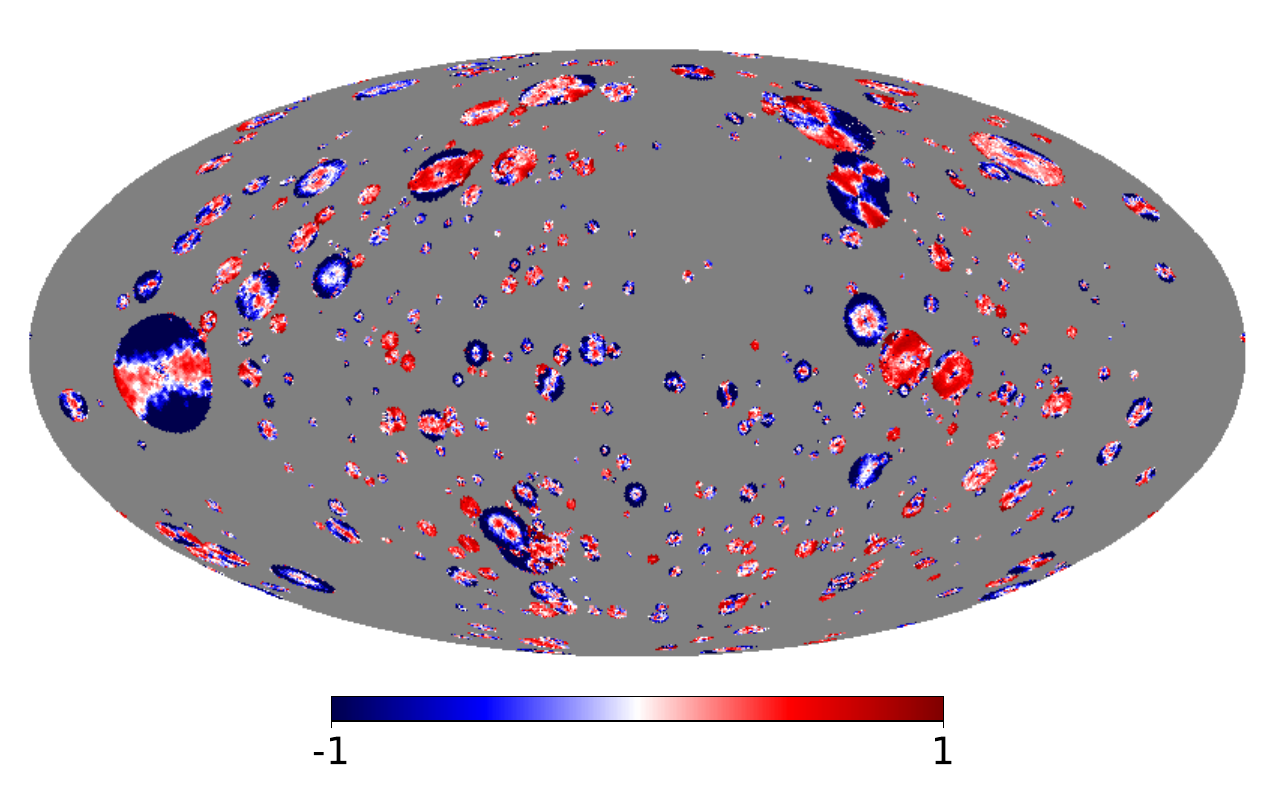}

    \caption{J-factor sky-map of the halo TNG-F computed with {\tt NAJADS} (top panel, left column) and map of the sub-structure contribution to the J-factor obtained subtracting the background contribution (bottom panel, left column). For comparison, the sub-halos contribution to the J-factor obtained integrating the analytical profiles is plotted in the top panel, right column. The residuals $(J^{\rm sh} - J^{\rm sh}_{\rm th})/J^{\rm sh}_{\rm th}$ are showed in the bottom right panel
    (pixels in which the theoretical map goes to zero are masked). Units for the J-factor maps are $M_{\odot}^2/{\rm kpc}^5$.}

    \label{fig:TNG50_results}
\end{figure}

We close this section with the analysis of sub-structures. We added to the smooth TNG-S all particles belonging to the sub-halos identified by the {\tt SUBFIND} halo finder that satisfied two criteria: (i) the total number of particles associated to the sub-structure is higher than $2 \cdot 10^3$  and (ii) the distance between the sub-halo and the main halo centres is lower than $210 \, \rm{kpc}$. The former condition is based on the results of Sec. \ref{sec:testing} and guarantees that {\tt NAJADS} will efficiently resolve the detailed structure of the sub-halo, while the latter ensures that all selected sub-halos fall in the J-factor integration volume. We ended up with $547$ sub-structures, with masses in the range $9 \cdot 10^6 - 5 \cdot 10^9 \, M_{\odot}$.
We use {\tt NAJADS} to compute the J-factor map of this full halo. The results are shown in the top-left panel of Fig.~\ref{fig:TNG50_results}, where we can appreciate that the background (main) halo presents indeed a prolate shape, slightly tilted towards left. The sub-halos appear to be very bright, and some of them clearly emerge above the smooth contribution.
In the bottom-left panel of Fig.~\ref{fig:TNG50_results} we subtract the numerical background map (smooth halo) to highlight the sub-structures contribution, and compare it with the corresponding semi-analytical map (right column, top panel). The last (bottom, right) map in Fig.~\ref{fig:TNG50_results} displays the residuals of the sub-halo maps $(J^{\rm sh} - J^{\rm sh}_{\rm th})/J^{\rm sh}_{\rm th}$, where we masked the pixels in which $J^{\rm sh}_{\rm th}$ is zero. 
We clearly see that {\tt NAJADS} recovers efficiently all sub-structures in our sample. However, as it was the case for the main halo, not all sub-halos are accurately described by a spherical profile. Some halos even have very elongated shapes and overall bigger extensions\footnote{For the calculation of the analytical maps, the size of sub-halos was chosen as $r_{\rm max} = 2.5 \, r_{\rm vir}$, with $r_{\rm vir}$ being the virial radius.} than their theoretical partners, giving rise once again to peculiar features in the residual map. This point is particularly interesting for studies addressing the detection of sub-halos as extended sources in the sky (see e.g.~\cite{DiMauro:2020uos}). Again, {\tt NAJADS} promises to be a very useful tool to deploy in these analyses, allowing for more precise estimates of sub-halo extensions from simulations.
As for the intensity of the signal, both semi-analytical calculations and {\tt NAJADS} estimates yield J-factors of the same order of magnitude, though in some cases the theoretical profile tends to overestimate the signal with respect to {\tt NAJADS} calculations; in other sub-halos the opposite occurs. In the biggest sub-halos, both boosting and mitigation of the signal occur in different regions of the structure. Once again, we attribute this variability to the peculiar mass distribution, which deviates from the spherical model.


\section{Summary and Conclusions}
\label{sec:conclusions}

The identification of DM structures in cosmological simulations is often crucial for many astrophysical applications, among which the determination of DM indirect detection signatures. The DM density spatial profile of the identified (sub-)halos is usually assumed to approximate an analytical distribution, which is determined by fitting said distribution inside the structures recovered in cosmological simulations.
Whenever the statistics of particles inside the structure is not enough to allow a solid fit, the free parameters of the analytical profiles are set through phenomenological relations connecting them to the properties recovered by the halo finders algorithms, such as the halo total mass and concentration.
This semi-analytical approach is very practical in analysing sub-halo populations properties, and it allows the extremely fast computation of astrophysical J-factors. However, it also carries its own limitations. Among these, the necessary reliance on density profiles which might not accurately match the actual dark matter distribution, and the difficulty in modelling theoretically the background boost factor.

In order to address these limitations, in this work we introduced {\tt NAJADS}, a self-contained framework for the numerical calculations of astrophysical J-factors for dark matter studies.
The default version of {\tt NAJADS} takes as input raw outputs from cosmological simulations, relies on a nearest-neighbour algorithm to estimate the local dark matter density, and proceeds to compute the J-factors along the desired LOS.

{\tt NAJADS} has first been developed on test-case mock simulations (the Dragon catalogue), that have been built to mirror three different analytical DM density profiles for the galactic-sized smooth halos. Each mock simulation additionally included a population of sub-halos. 
Taking advantage of our knowledge of the input parameters upon which the Dragon catalogue was built, we tested {\tt NAJADS} capability of correctly recovering the local density and the sub-structures, at a satisfying accuracy for the task at hand.
We then applied {\tt NAJADS} to the J-factor calculation, a measure of the DM squared density along the LOS. 
We computed the numerical J-factor on the Dragon halos and compared it with the results obtained integrating the injected analytical density profiles.
With respect to the semi-analytical predictions, we showed that {\tt NAJADS} automatically includes in the computation the boosting of the signal due to local density fluctuations. We found that for our mock halos - where the density noise comes only from the Monte Carlo randomization - this boosting is only a moderate effect, affecting the J-factor signal up to, on average, less than $4\%$ and getting as high as $25 \%$ only for a handful of LOS. However, this effect becomes much higher in actual cosmological simualtions, where the noise originates from a multitude of physical processes.

A proof-of-concept application of {\tt NAJADS} was conducted on a simulated halo from the TNG-1-Dark run in the TNG50 simulation of the IllustrisTNG suite. We ran {\tt NAJADS} on both the smooth component and the sub-structures of this halo, computing the J-factor maps. Simultaneously, we computed the corresponding semi-analytical maps integrating an Einasto density profile, fitted over the simulated halo and sub-halos. Both semi-analytical calculations and {\tt NAJADS} estimates yield J-factors of the same order of magnitude. However, this halo has a much more complex 3D mass distribution than our spherical Dragon halos. Local anisotropies and density scatter have much higher impact and this reflects on the numerical J-factor, which deviates from its semi-analytical counterpart of up to $>300 \%$, depending on the LOS. The morphology of such deviations is of course dependent on the specific halo matter distribution {\it and} on the observer location in the halo.
This highlights a non-negligible level of uncertainty in using analytical profiles as good descriptors of simulated halos to compute phenomenological predictions.
The numerical approach does not remove such error, but facilitates the systematic study of a high number of halos and crucially allows the full characterization of its impact. This could be of great help in characterising the foreground noise to gamma surveys expected from the Milky Way DM halo and for addressing extra-galactic background anisotropies.

A fine reconstruction of the DM distribution in big halos has a number of other astrophysical applications, from studies on the systematics of detecting sub-structures through weak lensing to a deeper understanding of the connection between DM anisotropies an baryon dynamics in galactic systems. These are all tangential applications to which {\tt NAJADS} could be adapted. We also found that {\tt NAJADS} efficiently recovers the sub-structures in the TNG halos. For the biggest structures, it is capable of telling deviations from sphericity and determining the sub-structure extension, together with their impact on the total J-factor. This would be extremely relevant for statistical studies on sub-halos indirect detection as extended sources. Further development of {\tt NAJADS} and its application to DM indirect searches and beyond is left for future works.

\appendix
\section{Monte Carlo realizations of galactic halos: the Dragon catalogue}
\label{app:mock_catalogue}

Before launching {\tt NAJADS} on realistic cosmological simulations, we produce for testing purposes Monte Carlo (MC) realizations of galactic halos that mimic simulation outputs.  We rely on the results of existing N-body simulations to set the prescriptions for the mass distribution of the main host halo, together with the number, space and mass distributions of the sub-halos population.
We create a complete MC mock catalogue of galactic-sized halos, that we dub the {\it Dragon} suite.
We focus only on reproducing the particles' positions (neglecting their velocities), as it is all the information {\tt NAJADS} needs to compute densities and J-factors. All halos in the Dragon suite are marked by the code 'Dr-' followed by a progressive number (from 1 to 9), as detailed below.

\subsection{The main halo}
\label{subapp:mock_catalogue_main_halo}

Our halos are 3D realizations of the analytical mass distributions of Eqs.~\eqref{eq:analytical_profile}. It comes in handy to use the total virial mass of the halo as a free parameter, from which -- for a given $r_{-2}$ -- one can compute the overall normalisation $\rho_{-2}$ by simply requiring that the volume integral of the mass profile over the whole halo yields back the corresponding mass. 
For all the main halos in the Dragon catalogue, we fixed $r_{-2} = 15.14$ kpc and assumed a $\Lambda$CDM cosmological model with critical density today $\rho_c \simeq 148 \, {\rm M_{\odot}/kpc^3}$. 
The integral of the density is then taken up to the virial radius $r_{200}$, defined as the radius enclosing an average mass density equal to 200 times $\rho_c$. $r_{200}$ encloses the mass $M_{200}$, which is the halo virial mass.

We set the total virial mass at $10^{12} \, M_{\odot}$ and require that $10^8$ particles are contained within the virial radius. This determines the resolution of our simulations, fixing the mass of each particle at $m_p = 10^4 \, M_{\odot}$. This resolution is close to those of current cosmological simulations (see e.g. Aquarius (\cite{Springel:2008cc}), Via Lactea II (\cite{Diemand:2008in}), EAGLE (\cite{Schaller:2015}), TNG50 (\cite{Pillepich:2019})) and it allows us to resolve a few thousands sub-halos, as we will discuss in the next subsection.

We then simulate all mock halos in our catalogue up to $r_{\rm max} = 300$ kpc. We notice that this radius is always higher than $r_{200}$ for our halos, thus the total mass of each halo is higher than the virial mass.

Tab.~\ref{tab:dragon_halos_parameters} summarizes the analytical profile normalization, total number of particles and total mass for each family of Dragon halos. Our complete catalogue contains 9 halos, respectively realizations of the Einasto (Eq.~\eqref{eq:analytical_profile_a}, Dr-1$\div$3), NFW (Eq.~\eqref{eq:analytical_profile_b}, Dr-4$\div$6) and Burkert (Eq.~\ref{eq:analytical_profile_c}, Dr-7$\div$9) profiles.\\

\begin{table}
	\centering
	\begin{tabular}{|c|c|c|c|c|}
		\hline
		Name & Profile & $\rho_{-2} [{\rm M_{\odot}/kpc^3}]$ & $N_{\rm tot}$ & $M_{\rm tot} [{\rm M_{\odot}}]$ \\
		\hline
		Dr-1$\div$3 & Einasto & $2.94 \cdot 10^{6}$ & $112435783$ & $1.12 \cdot 10^{12}$  \\
		\hline
		Dr-4$\div$6 & NFW & $3.14 \cdot 10^{6}$ & $114054823$ & $1.14 \cdot 10^{12}$  \\
		\hline
		Dr-7$\div$9 & Burkert & $2.89 \cdot 10^{6}$ & $113674381$ & $1.14 \cdot 10^{12}$  \\
		\hline
	\end{tabular}
	\caption{\label{tab:dragon_halos_parameters} Relevant parameters for the three families of Monte Carlo-simulated main halos. $\rho_{-2}$ is the density profile overall normalization, $N_{\rm tot}$ is the total number of particles simulated within the maximum radius $r_{\rm max} = 300$ kpc, and $M_{\rm tot}$ is the corresponding total mass.}
\end{table}

\begin{figure}
	\centering
	\includegraphics[width=1.\textwidth]{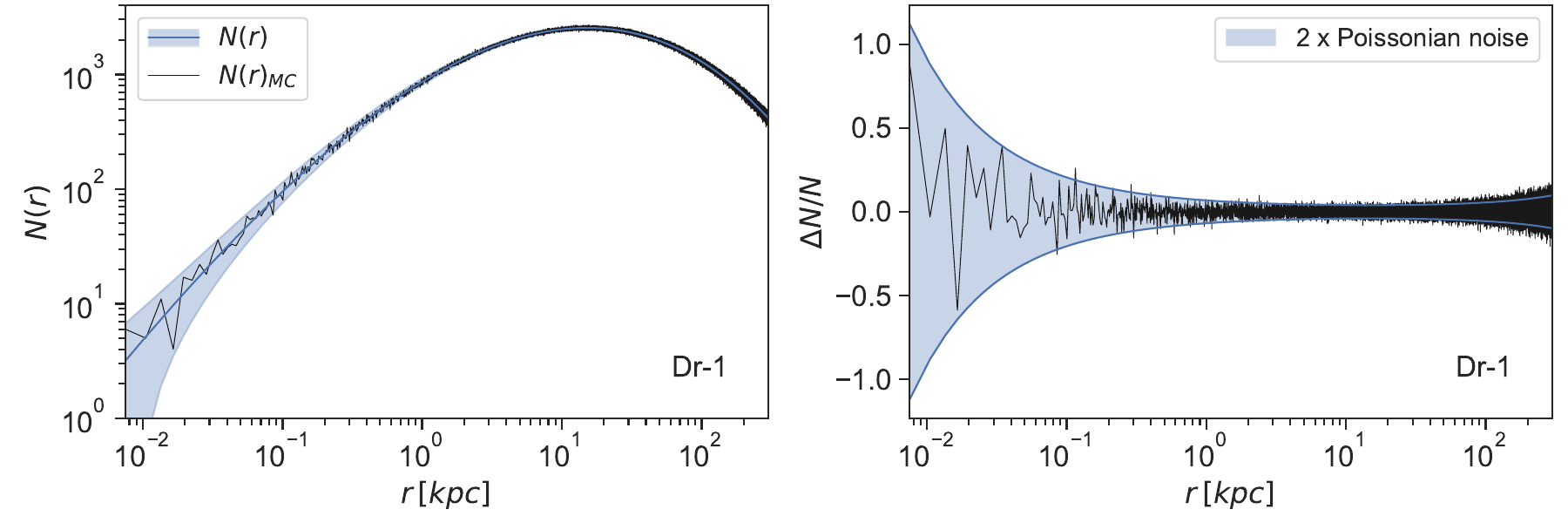}
	\caption{The left panel shows the theoretical number counts $N(r)$ vs the simulated one $N(r)_{\rm MC}$, for a smooth halo representative of an Einasto profile (Dr-1), as a function of the distance $r$ from the centre of the halo. In the right panel the relative mismatch between $N_{\rm MC}$ and $N$ is plotted for Dr-1. The grey-shaded area marks everywhere twice the expected Poissonian noise. 
	}
	\label{fig:number_count_halo}
\end{figure}

We simulate each mock halo using spherical coordinates, with the origin set at the halo centre. 
For each particle we drew the coordinates $\bar{x} = (r, \theta, \phi)$, where $\theta$ and $\phi$ are respectively latitude and longitude. The angular coordinates are randomly sampled from a uniform spherical distribution: $\phi$ is extracted in the interval $[0, 2\pi]$, while the latitude is computed as $\theta = \arccos (t)$, with $t \in [-1,1]$. This last device avoids the clustering of particles at the sphere poles.

Sampling the radial coordinate is more delicate. We start by computing the number of particles that should be present at any given radius to match the theoretical density profile. We dubbed this quantity $N(r)$.
More precisely,

\begin{equation}
\label{eq:disctretized_number_count}
\rho(r_i) = \frac{N(r_i, r_i+dr) \, m_{p}}{4\pi r_i^2 \, dr} \quad \rightarrow \quad \rho(r) = \lim\limits_{dr \rightarrow 0} \frac{N(r) \, m_{p}}{4\pi r^2 \, dr} \, ,
\end{equation}
where the first equation refers to a discrete realisation of the continuous mass distribution $\rho(r)$. $N(r_i, r_i+dr)$ is then the number of particles of mass $m_p$ within a spherical shell of radius $r_i$ and thickness $dr$. Hence, we proceed to divide the halo into nested spherical shells, ideally thin enough to resemble as much as possible a continuous distribution, but still thick enough that $N(r_i) \geq 1$ for all $i$. The particles' radial coordinates are then sampled from the $N(r_i)$ discretised distribution. This process yields a new distribution $N(r_i)_{\rm MC}$ which matches the initial one barring some small fluctuations in the form of a poissonian noise $\Delta N(r_i) = \sqrt{N(r_i)}$. We further smooth out our simulations by re-drawing the radius of all $N(r_i)_{\rm MC}$ particles in the $i^{\rm th}$ shell from the interval $[r_i-dr/2, r_i+dr/2]$.

In Fig.~\ref{fig:number_count_halo} we plot the comparison between $N(r_i)$ and $N(r_i)_{\rm MC}$ for the Dr-1 halo. The two profiles are in good agreement, the biggest discrepancies occurring at the outskirts and in the centre below approximately $r = 0.1$ kpc, where the number of particles is smaller. We analysed the same plot for all our MC halos, and found that these differences are slightly higher for Burkert halos, as such profiles exhibit a lower central density with respect to the cuspy ones, predicting even less particles in the central regions. However, all mismatches are contained within twice the predicted poissonian noise, hence we are satisfied that our MC halos are accurate enough realizations of the injected mass distributions.

\subsection{The sub-halos}
\label{subapp:mock_catalogue_subhalos}

We model the statistical properties of our sub-halos populations on the sub-structures found in the second run of the Via Lactea II (hereafter VLII) simulation suite~\citep{Diemand:2008in}, exploiting the data of their public catalogue\footnote{\href{https://www.ucolick.org/~diemand/vl/index.html}{https://www.ucolick.org/~diemand/vl/index.html}}. 
Halos and sub-halos in VLII were identified with the {\tt 6DFOF} halo finder \citep{Diemand:2006ey}. Of all 20048 sub-structures present in the catalogue,
we select the 3472 ones that are located within a radius $10 \leq r \leq 250 \, {\rm kpc}$ from the main halo centre and have masses higher than $10^6 \, {\rm M_{\odot}}$, i.e. containing at least 100 particles at the resolution of our MC halos.

\begin{figure}
	\includegraphics[width=0.328\textwidth]{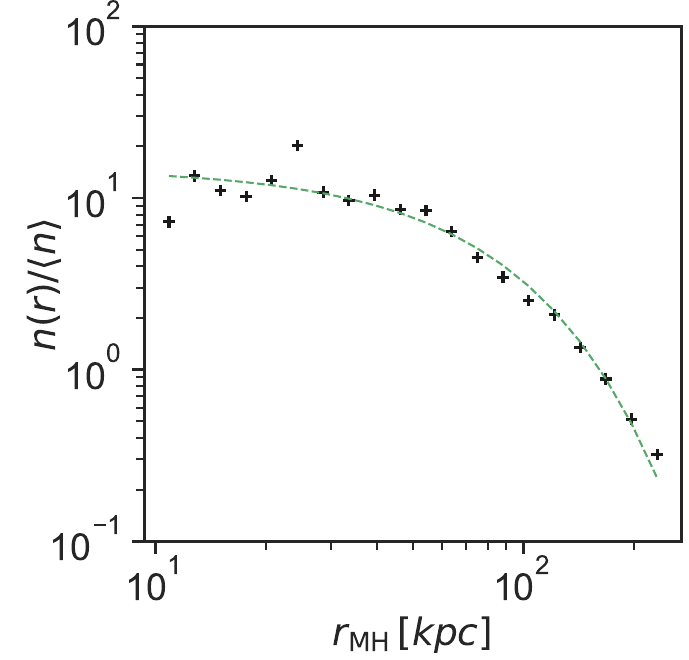}
	\includegraphics[width=0.328\textwidth]{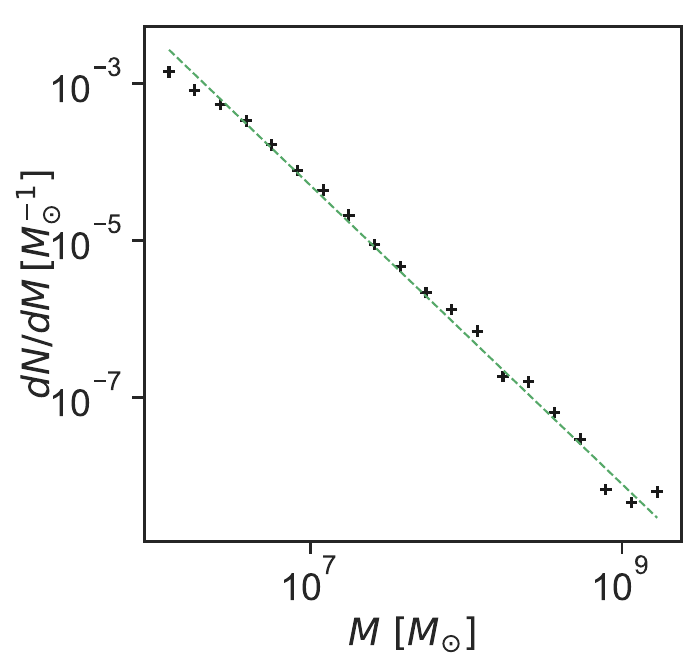}
    \includegraphics[width=0.328\textwidth]{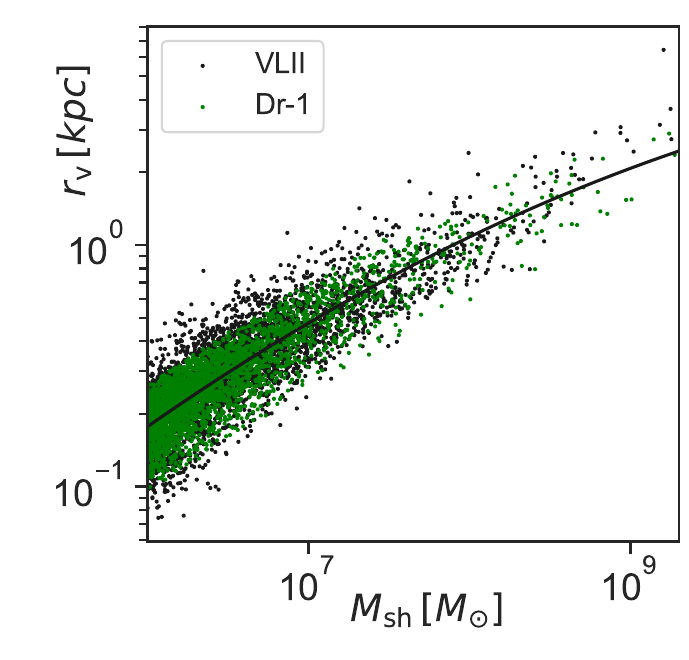}
    \caption{Fits of the radial number density (left panel) and differential mass abundance (middle panel) of sub-halos on the reduced catalogue from VLII. $r_{\rm MH}$ refers to the position of the sub-halo with respect to the main halo centre. The right panel displays the scatter plot of the mock sub-halos population of the Dr-1  halo in the $r_{\rm v} - M_{\rm sh}$ plane, compared to the subhalo population of VLII. The solid line shows the fit of the phenomenological correlation $r_{\rm v} - M_{\rm sh}$ to the sub-structures from the reduced VLII catalogue.}
	\label{fig:subhalo_fits}
\end{figure}

The spatial distribution of sub-halos can be interpolated with a radial function, whose parametrization resembles that of an Einasto profile (see e.g. \citep{Zhu:2015jwa} and \citep{Calore:2016ogv}):
\begin{equation}
\frac {n(r_{\rm MH})}{n_{\rm tot}} = n_{-2} \cdot \exp\left\{ -\frac{2}{\alpha_n} \Big[\Big(\frac{r_{\rm MH}}{r_{-2}}\Big)^{\alpha_n} -1 \Big]\right\} \, ,
\label{eq:subhalos_fit_count}
\end{equation}
where $r_{\rm MH}$ is the distance from the main halo centre, $n(r_{\rm MH})$ is the sub-halos number density at given radial bin and $n_{\rm tot}$ is the total number of structures in the sample. The parameters $\alpha_n\text{, }n_{-2}\text{ and }r_{-2}$ are fitted against the reduced VLII catalogue, yeilding $\alpha_n = 1.21 \pm 0.16$, $n_{-2} = 2.82 \pm 0.04$ and $r_{-2} = (108 \pm 2) \, $kpc. 
The results are shown in the left panel of Fig.~\ref{fig:subhalo_fits}.

The mass distribution of sub-structures is also well know (see for example \citep{Zhu:2015jwa, Calore:2016ogv, Springel:2008cc}). These authors all found that the differential mass abundance of sub-halos in cosmological simulations is well reproduced by a power-law of the form
\begin{equation}
\frac{dN}{dM} = M_0 \Big(\frac{M}{M_{\odot}}\Big)^{-\alpha_{\rm M}} \, .
\label{eq:subhalos_fit_masses}
\end{equation}
Eq.~\eqref{eq:subhalos_fit_masses} is also fitted on the reduced VLII catalogue, and the results are plotted in the middle panel of Fig.~\ref{fig:subhalo_fits}. We obtain $M_0 = (0.98 \pm 0.22) \cdot 10^9 \,  {\rm M_{\odot}^{-1}}$ and $\alpha_{\rm M} = 1.9 \pm 0.1$, this last value in particular also in agreement with those found in \citep{Springel:2008cc} and \citep{Zhu:2015jwa} for the halos of the Aquarius suite.

Moreover, \cite{Calore:2016ogv} found for the C-halo of Aquarius a correlation between the sub-halo masses and the radius at which the structure's particle velocities peak, $r_{\rm v}$. In addition, \cite{Springel:2008cc} showed that there is a phenomenological relation connecting $r_{\rm v}$ to the halo scale radius: $r_{\rm v} \simeq 2.189 \, r_{-2}$.
The combination of these results gives

\begin{equation}
\log_{10} \Big(\frac{r_{-2}}{\rm kpc}\Big) = p_0 - \log(2.189) + p_1 \log_{10}\Big(\frac{M_{ \rm sh}}{M_{\odot}}\Big) + p_2 \Big(\log_{10}\Big(\frac{M_{\rm sh}}{M_{\odot}}\Big)\Big)^2 \, .
\label{eq:subhalos_fit_scale_radius}
\end{equation}

Once again, we fit Eq.~\eqref{eq:subhalos_fit_scale_radius} on the sample of sub-halos from VLII and verify that this correlation still holds. We obtain $p_0 = -4.840$, $p_1 = 0.898$ and $p_2 = -0.037$. We additionally include in our model the population scatter around the best fit through the relative standard deviation $\sigma_{r_{-2}}/r_{-2} = 0.237$.
The results are presented in the right panel of Fig. \ref{fig:subhalo_fits}.\\

\begin{figure}
	\centering
	\includegraphics[width=1.\textwidth]{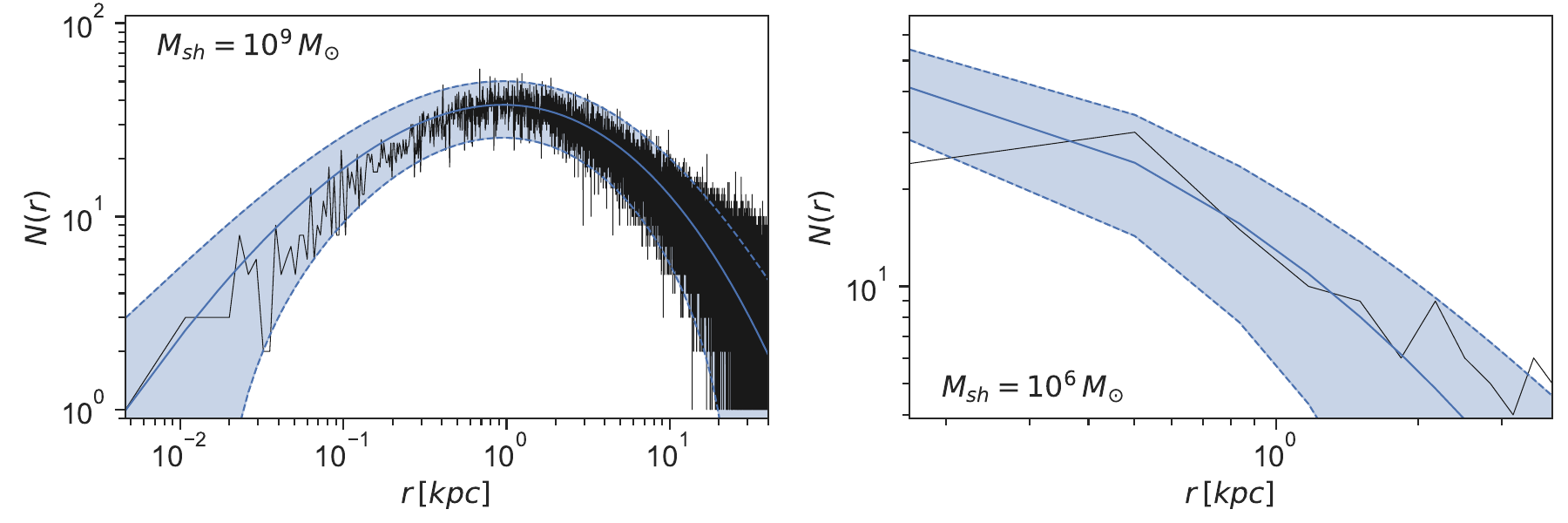}
	\caption{Theoretical number counts $N(r)$ (blue line) compared to the simulated one $N(r)_{\rm MC}$ (black line) for sub-halos bracketing the explored mass range. The shaded grey area marks twice the poissonian error.}
	\label{fig:number_count_subhalos}
\end{figure}

With all the ingredients in place, we add a population of sub-structures to every mock Dragon halo, following three steps: i) the mass is sampled from the distribution \eqref{eq:subhalos_fit_masses} in the interval $10^5  \leq M_{\rm sh} \leq 10^9 \, {\rm M_{\odot}}$ and $r_{-2}$ is computed using \eqref{eq:subhalos_fit_scale_radius} and accounting for scattering; ii) the structure is placed in the main halo by sampling Eq.~\eqref{eq:subhalos_fit_count}; iii) each sub-halo is simulated as described in Sec.~\ref{subapp:mock_catalogue_main_halo}, reflecting the same mass profile of its host with normalisation $\rho_{-2}$ determined from its total mass.

We restricted to sub-halos with masses higher than $10^5 ~{\rm M_{\odot}}$ so that all Dragon sub-structures contain between $10$ and $10^5$ particles. 
In Fig.~\ref{fig:number_count_subhalos} we plot the particle number counts $N_{\rm MC}(r_i)$ compared to the theoretical $N(r)$ for two examples of MC sub-halos with mass $M_{sh} = 10^9 \, {\rm M_{\odot}}$ and $M_{sh} = 10^6 \, {\rm M_{\odot}}$ respectively. As sub-halos contain way fewer particles than their host halos, the numerical poissonian noise has a higher impact, so mismatches between $N_{\rm MC}(r)$ and $N(r)$ are more relevant -- especially in the inner regions where the number of particles is lowest. In particular for sub-halos of masses $10^6 \, {\rm M_{\odot}}$ (and lower), that contain a total of only 100 particles, a fine reproduction of the input $N(r)$ is very difficult.

\section{Density scatter in MC simulated halos}
\label{app:density_scatter}

We highlighted in Sec.~\ref{subsec:density_tests} that local numerical estimates of the density at any given radius in our mock MC halos present small variations from the expected average value of the density at that same radius. These local fluctuations are physiological of simulated halos, and it is then interesting to characterize the dispersion $\sigma$ of the density around the average value at different radii, and gauge its impact on our conclusions.

\begin{figure}
	\includegraphics[width=0.329\textwidth]{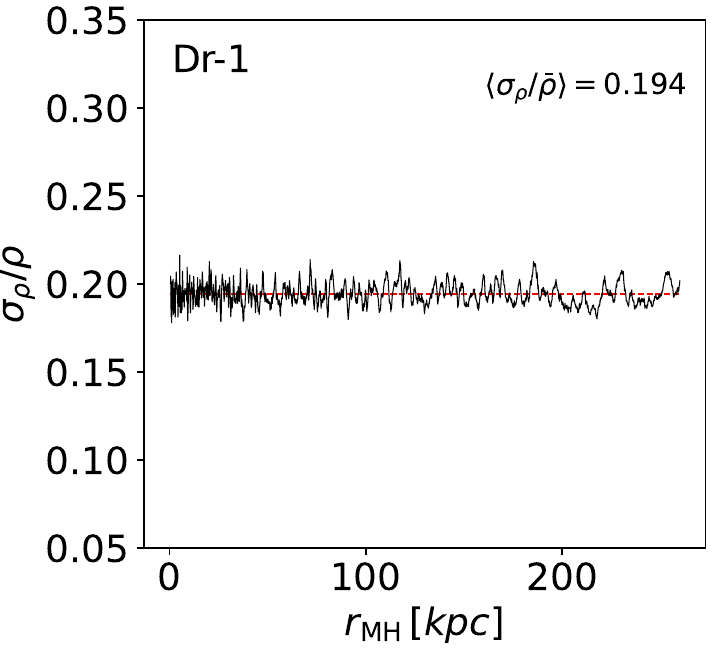}
	\includegraphics[width=0.329\textwidth]{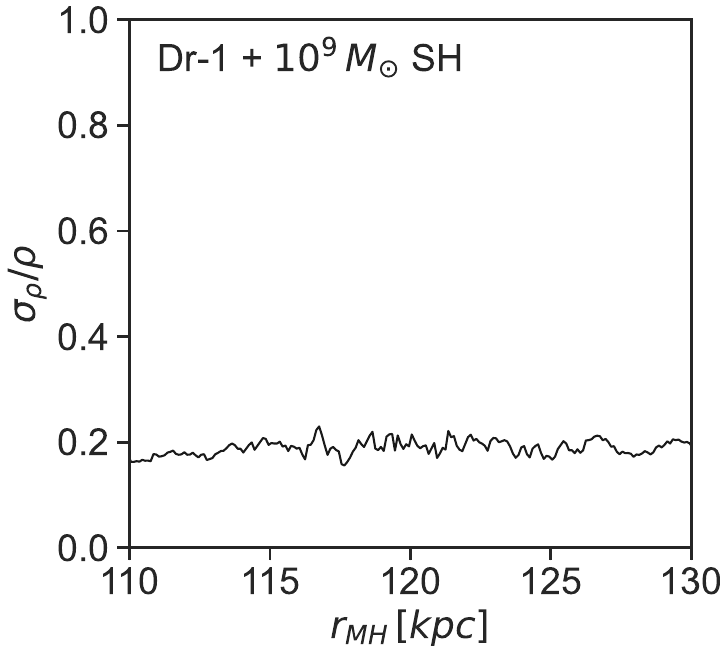}
	\includegraphics[width=0.329\textwidth]{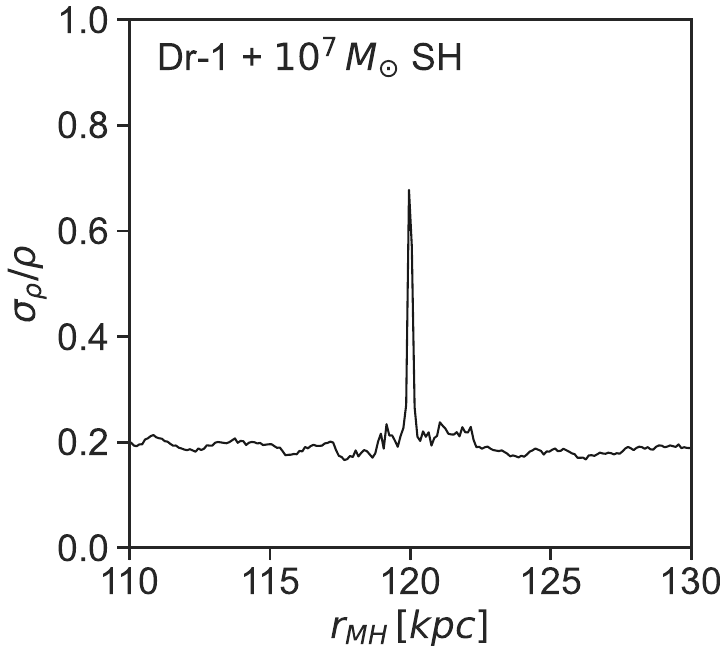}

	\caption{Relative density dispersion profiles $\sigma_{\rho}/\rho$ computed with {\tt NAJADS} for halos representative of an Einasto density profile, as a function of the distance from the centre of the main halo $r_{\rm MH}$. The three plots are obtained, respectively, for the case of a smooth halo (left panel) and inserting a sub-halo of mass $10^9 \, M_{\odot}$ (middle panel) or $10^7 \, M_{\odot}$ (right panel) along the LOS. Middle and right panels are zoomed-in on $r_{\rm MH}$ at the position of the sub-halo.}
	\label{fig:sigma_halo+subhalos_E}
\end{figure}

In the left panel of Fig.~\ref{fig:sigma_halo+subhalos_E} we plot the ratio $\sigma_{\rho}/\rho$ as a function of radius $r_{\rm MH}$ for the halo Dr-1. We observe that the relative density dispersion settles around the same flat value of $\sim 0.194$ over the whole halo, with no other visible trend aside from small oscillations. We find similar results for all Dragon halos, regardless of the theoretical mass distribution used as input for their MC realisation. This is reassuring, as all sections of the halo were simulated using the same procedure, so there is no reason why different sections should be affected differently by fluctuations.
However, the fact that we are able to spot fluctuations \textit{per se} already speaks about the advantages of introducing a tool for the J-factor calculation that computes the DM density point-to-point in the simulation instead of relying on averaged profiles.

In our mock halos, we find that the local density can present deviations of $\sim 20\%$ from the average value. These fluctuations are conceptually similar to those found in halos in realistic cosmological simulations.
The latter will, however, differ quantitatively, as they originate from a number of distinct physical effects. These may include deviations from spherical symmetry of the main halo mass distribution, residuals of disrupted sub-structures, structures resulting from previous mergers with other halos and not yet fully smoothed out, or an unresolved population of sub-halos (see e.g. \citep{Zavala:2019gpq} for a review). Each of these factors impacts the relative mass dispersion in various regions of the halos in different ways. Thus, in cosmological halos, $\sigma_{\rho}/\rho$ can in principle reach much higher values than $20 \%$, and exhibit the most disparate trends, possibly encapsulating important physical information on the halo's formation history. In this sense, we caution that the results of Fig.~\ref{fig:sigma_halo+subhalos_E} apply solely to our MC halos and are not to be interpreted as universal.\\

We move to characterise the changes in the density dispersion when a substructure is present. Similarly to what was done in Sec.~\ref{subsec:density_tests}, we re-compute the dispersion profiles $\sigma_{\rho}(r)/\rho(r)$, considering all LOS within a given halo, pierced by the MC realization of a sub-halo of fixed mass. The results are shown in the central and right panels of Fig.~\ref{fig:sigma_halo+subhalos_E}. For bigger sub-halos ($10^9- 10^8 \, M_{\odot}$), we find that the flat behaviour that $\sigma_{\rho}/\rho$ exhibits for the main halo does not change.
However, a spike starts to appear in the dispersion profiles when the sub-structure mass drops to $10^7 \, M_{\odot}$ and lower, as shown in the right panel of Fig.~\ref{fig:sigma_halo+subhalos_E} for the Dr-1 halo. This spike indicates that the density scatter increases significantly in the region corresponding to the centre of the sub-halo.

We verify this finding with a one-by-one analysis of the density profiles computed for all LOS piercing a medium mass halo ($\sim 10^7 \, M_{\odot}$), which highlighted that in a number of cases {\tt NAJADS} estimates densities in the central regions of the structure that deviate significantly from the injected analytical profiles. The code either underestimates or overestimates the expected signal, depending on the specific halo. This behaviour can, in good measure, be linked to the MC procedure pitfalls in halo centres. 

Essentially, in the very small volumes that correspond to the inner regions of sub-halos of smaller masses, only a tiny number of particles are necessary - at our resolution - to reproduce the injected theoretical density. As particles positions are randomly drawn during the MC process, drawing a couple extra particle in that region can introduce significant differences in the simulated densities. Indeed, scanning all small sub-halos simulated for this test, we find a correlation between {\tt NAJADS} under(over)-estimates of the central density peak and the respective discrepancy between the simulated and expected number of particles in the sub-halos cores.
In other words, the scatter in the density distributions of Fig.~\ref{fig:sigma_halo+subhalos_E} is higher in the same regions for which the poissonian noise associated with the MC sampling has a bigger impact.
Nonetheless, this does not mean that {\tt NAJADS} is failing in computing the densities for these halos. On the contrary, the average numerical $\rho(r)$ still matches very well the injected profile -- where the average is performed over all $10^3$ LOS, and hence $10^3$ different realisations of same-mass sub-halos. This means that the increase in variance is not accompanied by a systematic under(over)-estimate of the density, which would instead be a clear sign of {\tt NAJADS} poor performance.\\

\section{Determination of (a,b) parameters}
\label{app:ab_determination}

We explain here how we determine the best choice for the $(a, b)$ couple of free parameters. We consider that, taken a particle in the simulation, there are 6 main spatial directions along which nearest neighbours can be looked for: up, down, forward, backward, right and left. Along each of these directions, it is reasonable to expect either 1 or 2 particles to be actual nearest neighbours. This empirical reasoning brought us to speculate values for $a$ and $b$ respectively of 6 and 12. We then considered the configurations $(a,b) = (5,13), \, (6,13)\, (6,12), \, (7,12), \, (7,11), \, (8,11), \, (8,10), \, (9,10)$, all having in common an average number of nearest neighbours of $\simeq$ 9. We recall that the principal constraint we imposed when generating the mock simulations is that the resulting halos should present a specific radial density distribution $\rho(r)$. Thus, the best parameters couple $(a,b)$ can be identified based on which choice is actually able to better reconstruct $\rho(r)$ for any given halo.
We ran {\tt HADES} on each of the 9 smooth halos\footnote{i.e. the MC host halos \textit{before} the addition of the sub-structures.} in the Dragon catalogue, selecting in turn each of the 8 $(a,b)$ configurations listed above. More specifically, for each halo we evaluated the density with all chosen {\tt HADES} configurations at 1000 points equally distant from the halo centre, and repeated the procedure at 10 progressive radii, for a total of $10.000$ estimation spots.

\begin{figure*}
	\centering
	\includegraphics[width=\textwidth]{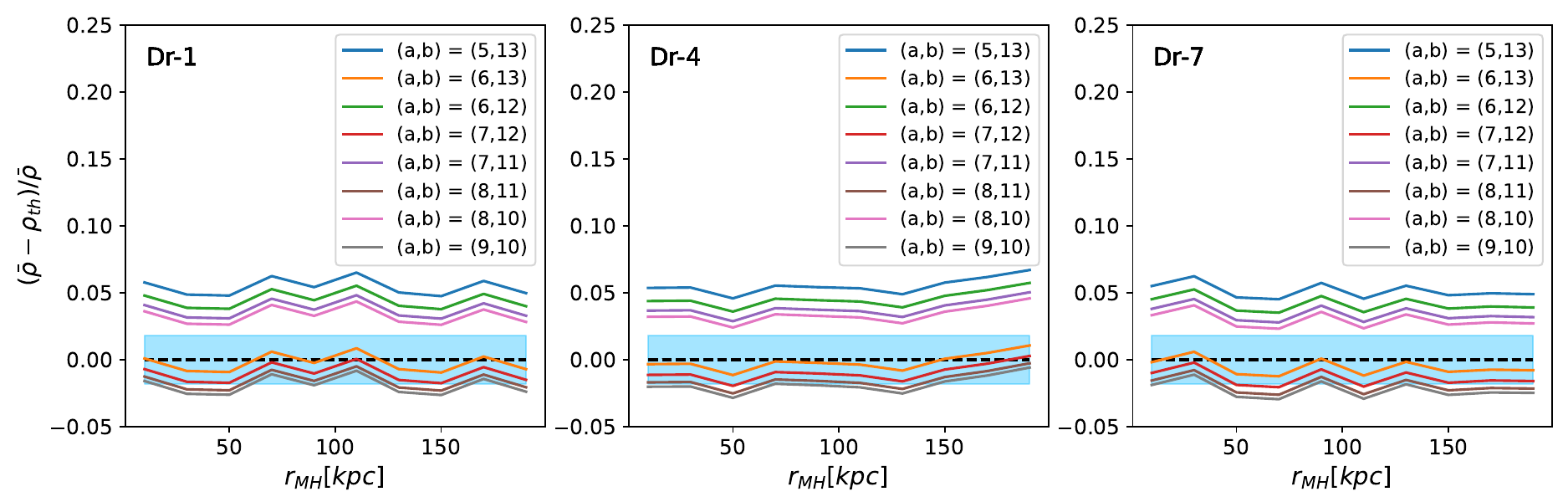}
	\caption{ Relative difference between the theoretical density $\rho_{\rm th} = \rho(r_{\rm MH})$ and the average density $\bar{\rho}$ computed with a sample of 1000 measures at 10 different radii $r_{\rm MH}$ inside an MC simulated halo. Each colour line corresponds to a different choice of {\tt NAJADS} parameters $(a,b)$. The three panels show the results for three different MC halos in the Dragon catalogue, Dr-1, 4 and 7.}
	\label{fig:ab_couple}
\end{figure*}

In Fig. \ref{fig:ab_couple}, we plot the ratio $\delta \rho/\bar{\rho}$ at all radii $r_{\rm MH}$ we considered in our test, for three of the Dragon halos. The light blue area marks the $3\sigma$ relative error over the expectation value of the density distribution. The fluctuations of all these lines around the horizontal 0 value suggest that the true density distribution of the halos itself slightly differs from $\rho_{\rm th}(r)$. We recall that the MC halos reproduce the theoretical $\rho_{\rm th}(r)$ only within a given error, hence small deviations are expected. This scenario is supported by the accordant trend of the curves of Fig. \ref{fig:ab_couple}, each one seeming to over(under)-estimate the same underlying density of a fixed amount.
From Fig. \ref{fig:ab_couple} we can clearly see that the couples $(5,13)$, $(6,12)$, $(7,11)$ and $(8,10)$ can be set aside, as they introduce a systematic overestimate of the density that is higher than the $3\sigma$ uncertainty. On the contrary, the remaining 4 couples $(6,13)$, $(7,12)$, $(8,11)$ and $(9,10)$ seem to all be acceptable candidates, each of them slightly underestimating the average density of less than $\sim 2 \%$. However, the couple $(6,13)$ evidently yields the better result, always deviating from 0 of at most $1\%$. 
This same result holds for all the halos in the Dragon catalogue, hinting that it is generally valid for any kind of density profiles $\rho(r)$ typically found in DM halos. This is reassuring, as we wish to chose a parameter configuration to run {\tt NAJADS} on simulations for which we have no control over the real underlying DM density, ideally introducing the smallest possible error.

To further check that the choice $(6,13)$ is indeed the best independently of the density profile, we repeated the test performed in this Section on a number of MC simulations following very different mass distributions, from a uniform realisation, to power-laws of different steepness, to density gradients spanning up to 5 orders of magnitude.
In each case, the couple $(6,13)$ always resulted the best possible parameters choice, going as far as to reconstruct the average density with a precision of $0.04\%$ in the case of a uniform realisation.  This makes us reasonably confident that {\tt HADES} can be applied safely to real simulations as well, with the suggested choice $(a, b) = (6, 13)$ possibly introducing only systematic reaching in the worst case scenario the $\%$ level. We further consolidate our choice for the algorithm parameters in Sec.~\ref{sec:testing}, where we show that we can apply it to mock sub-halos of various sizes, robustly recovering the injected density profile.

The same mock simulations used to fix the free parameters of our density module were exploited for a number of numerical tests, among which the check that not only the average density, but also the total volume of the simulation is correctly recovered. Specifically, we used our code to estimate the density (and thus, the infinitesimal volume) at each particle position. Summing all infinitesimal volumes calculated in this way, we were in all cases able to recover the total volume of the mock simulation within a maximum 10\% error.

\section*{Acknowledgments}

We thank Dylan Nelson for giving us access to the IllustrisTNG simulations data and Moritz H\"utten for helping us with the installation of {\tt CLUMPY}.
AB is supported by a de Sitter Fellowship of the Netherlands Organization for Scientific Research (NWO).
VDR acknowledges financial support by the CIDEXG/2022/20 grant (project “D’AMAGAT”) funded by Generalitat Valenciana, and by the Spanish grant PID2020-113775GB-I00 (MCIN/AEI/10.13039/ 501100011033).
The work of FD is  supported by the {\sc Departments of Excellence} grant awarded by the Italian Ministry of Education,
University and Research ({\sc Miur}) and by the 
Research grant {\sc TAsP} (Theoretical Astroparticle Physics) funded by Istituto
Nazionale di Fisica Nucleare.

\bibliographystyle{utphys}
\bibliography{bibliography}

\end{document}